\documentclass[english,10pt]{article}

\usepackage[english]{babel}
\usepackage[utf8]{inputenc}
\usepackage[T1]{fontenc}

\usepackage{geometry}
\usepackage{graphicx}
\usepackage[colorinlistoftodos]{todonotes}
\usepackage[colorlinks=true, allcolors=blue]{hyperref}
\usepackage{setspace}
\usepackage{enumitem}
\usepackage{titlesec}
\usepackage{comment}
\usepackage{caption}
\usepackage{subcaption}
\usepackage{longtable}
\usepackage{booktabs}

\usepackage[section]{placeins}

\usepackage{cite}

\usepackage{amsthm}
\usepackage{amssymb}
\usepackage{amsfonts}
\usepackage{bbm}
\usepackage{bm}
\usepackage{nicefrac}
\usepackage{mathtools}
\usepackage{amsmath}

\usepackage{subfiles}

\usepackage[page]{appendix}

\usepackage{fullpage}
\usepackage{float}
\usepackage{wrapfig}

\usepackage[boxruled]{algorithm2e}
\usepackage{listings}
\usepackage{dsfont}
\usepackage{array}
\usepackage{bbold}

\makeatother

\usepackage{color}
\usepackage{centernot}

\definecolor{ejc}{RGB}{255,0,0}
\definecolor{ms}{RGB}{0,100,0}
\definecolor{yr}{RGB}{0,100,100}

\newcommand{\indep}{\rotatebox[origin=c]{90}{$\models$}}

\DeclareMathOperator*{\argmin}{arg\,min}

\newcommand{\E}[1]{\mathbb{E}\left[#1\right]}

\newcommand{\Ec}[2]{\mathbb{E}\left[\left.#1\right|#2\right]}

\newtheorem{theorem}{Theorem}

\newcommand{\reals}{\mathbb{R}}

\newcommand{\Normal}[2]{\mathcal{N}\left(#1,#2\right)}

\allowdisplaybreaks[1]

\RequirePackage{csquotes}

\definecolor{mygreen}{rgb}{0,0.6,0}
\definecolor{codegray}{rgb}{0.98,0.98,0.98}
\definecolor{mauve}{rgb}{0.58,0,0.82}
\def\X{{\mathbf X}}

\def\Y{{\mathbf Y}}

\def\Z{{\mathbf Z}}

\def\RR{{\mathbb R}}
\def\EE{{\mathbb E}}

\def\V{{\mathbf V}}

\def\pperp{{\ \perp\kern-10pt\perp \ }}

\newlength\mylen
\newcommand\myinput[1]{%
	\settowidth\mylen{\KwIn{}}%
	\setlength\hangindent{\mylen}%
	\hspace*{\mylen}#1\\}

\title{Deep Knockoffs}
\author{Yaniv Romano\thanks{These authors are listed in alphabetical order.} \thanks{Department of Statistics, Stanford University, Stanford, CA 94305, U.S.A.} ,
Matteo Sesia\footnotemark[1] \footnotemark[2],
Emmanuel J. Cand\`es\footnotemark[2] \thanks{Department of Mathematics, Stanford University, Stanford, CA 94305, U.S.A.}}

\begin{document}
\maketitle

\begin{abstract}
This paper introduces a machine for sampling approximate model-X
knockoffs for arbitrary and unspecified data distributions using deep
generative models.  The main idea is to iteratively refine a knockoff
sampling mechanism until a criterion measuring the validity of the
produced knockoffs is optimized; this criterion is inspired by the
popular maximum mean discrepancy in machine learning and can be
thought of as measuring the distance to pairwise exchangeability
between original and knockoff features.  By building upon the existing
model-X framework, we thus obtain a flexible and {\em model-free}
statistical tool to perform controlled variable selection. Extensive
numerical experiments and quantitative tests confirm the generality,
effectiveness, and power of our deep knockoff machines. Finally, we
apply this new method to a real study of mutations linked to changes
in drug resistance in the human immunodeficiency virus.



\end{abstract}

\section{Introduction} \label{sec:introduction}
\subsection{Motivation}


Model-X knockoffs \cite{candes2016panning} is a new statistical tool
that allows the scientist to investigate the relationship between a
response of interest and hundreds or thousands of explanatory
variables. In particular, model-X knockoffs can be used to identify a
subset of important variables from a larger pool that could
potentially explain a phenomenon under study while rigorously
controlling the false discovery rate \cite{benjamini1995controlling}
in very complex statistical models. While this methodology does not
require any knowledge of how the response depends on the values of the
features, the correctness of the inferences rests entirely on a
precise description of the distribution of the explanatory variables,
which are assumed to be random. This makes model-X knockoffs well
adapted to situations in which good models are available to describe
the joint distribution of the features, as in genome-wide association
studies \cite{sesia2017gene} where hidden Markov models are widely
used to describe patterns of genetic variation. To apply the knockoffs
approach in a broad set of applications, however, we would need
flexible tools to construct knockoff variables from sampled data in
settings where we do not have reliable prior knowledge about the
distribution of the covariates but perhaps sufficiently many labeled
or unlabeled samples to `learn' this distribution to a suitable level
of approximation.  These conditions are realistic because the
construction of model-X knockoffs only depends on the explanatory
variables whose unsupervised observations may be abundant. For
example, even though the genome-wide association analysis of a rare
disease may contain a relatively small number of subjects, the genetic
variants for other individuals belonging to the same population can be
gathered from different studies.

The goal of this paper is simply stated: to extend the applicability
of the knockoffs framework as to make it practically model-free and,
therefore, widely applicable.
This is achieved by taking advantage of important recent progress in
machine learning, which is repurposed to harness the information
contained in large unsupervised datasets to sample approximate model-X
knockoffs. The ultimate outcome is a set of sensible and flexible
tools for model-free controlled variable selection that can help
alleviate the crucial irreproducibility issues afflicting many areas
of science and data analysis \cite{ioannidis2005most, gelman2014statistical, baker20161, munafo2017manifesto}. 
A preview of our contribution is sketched below, while the technical details are
postponed to later sections.

\subsection{Our contribution}

Given independent copies of $X = (X_1, \ldots, X_p) \in \RR^p$ from
some unknown distribution $P_X$, we seek to construct a random
generator of valid knockoffs
$\tilde{X} = (\tilde{X}_1, \ldots, \tilde{X}_p)$ such that the joint
law of $(X,\tilde{X})$ is invariant under the swapping of any $X_j$
and $\tilde{X}_j$ for each $j \in \{1,\ldots,p\}$ (see Section
\ref{sec:knockoffs} for details).  Concretely, the machine takes the
data $X$ as input and generates $\tilde{X}$ through a mapping
$f_\theta(X, V)$, where $V$ is random noise, and $f_\theta$ is
a deep neural network. The parameters of the network are fitted on
multiple observations of $X$ to optimize a loss function that
quantifies the extent to which $\tilde{X}$ is a good knockoff copy of
$X$. This goal is related to the classical problem of learning
generative models; however, the challenge here is unusual since only
$X$ is accessible while no sample from the target distribution
$P_{\tilde{X} \mid X}$ is available. Fortunately, the existing methods
of deep generative modeling reviewed in Section \ref{sec:deep-models}
can be suitably repurposed, as we shall see in Section
\ref{sec:machines}.  Furthermore, the lack of uniqueness of the target
distribution raises an additional question. Intuitively, this
ambiguity should be resolved by making $\tilde{X}$ as different as
possible from $X$, since a trivial copy---setting
$\tilde{X} = X$---would satisfy the required symmetry without being of
any practical use for variable selection. Our approach generalizes the
solution described in \cite{candes2016panning}, which relies on the
simplifying assumption that $X$ can be well-described as a
multivariate Gaussian vector. In the context of deep generative
models, the analogous idea consists of training a machine that
optimizes the compatibility of the first two moments of
$(X,\tilde{X})$ while keeping the strength of the pairwise
correlations between $X_j$ and $\tilde{X}_j$ for each
$j \in \{1,\ldots,p\}$ under control. By including in the loss
function an additional term that promotes the matching of higher
moments, we will show that one can move beyond the second-order
approximation towards a model-free knockoff generator.  The
effectiveness of deep knockoff machines can be quantitatively measured
using the goodness-of-fit diagnostics presented in Section
\ref{sec:robust-diagnostics}, as shown empirically by the results of
our numerical experiments (Section \ref{sec:experiments}) and data
analysis (Section \ref{sec:application}).  The algorithms described in
this paper have been implemented in Python and the corresponding
software is available from
\url{https://web.stanford.edu/group/candes/deep-knockoffs/}.

\subsection{Related work}

The main idea of using knockoffs as negative control variables
was originally devised in the context of 
linear regression setting with a fixed design matrix \cite{barber2015}.
The generation of model-X knockoffs beyond the settings considered in
\cite{candes2016panning} and \cite{sesia2017gene} has also been
tackled in \cite{gimenez2018knockoffs}, which extends the results for
hidden Markov models to a broader class of Bayesian networks. More
recent advances in the framework of knockoffs include the work of
\cite{lu2018deeppink, fan2018ipad, zheng2018}, while some interesting
applications can be found in \cite{xiao2017mapping, xie2018false,
  Gao2018}.  Very recently, deep generative models have independently
been suggested as a procedure for sampling knockoffs in
\cite{anonymous2019knockoffgan}; there, the approach focuses on
adversarial rather than moment matching networks. Even though the
fundamental aims coincide and the solutions are related, our machine
differs profoundly by design and it offers a more direct connection with existing work
on second-order knockoffs. Also, it is well known that generative
adversarial networks are difficult to train
\cite{arjovsky2017towards}, while moment matching is a simpler task
\cite{li2015generative, dziugaite2015training}. Since the approach of
\cite{anonymous2019knockoffgan} requires simultaneously training four
different and interacting neural networks, we expect that a good
configuration for our machine should be faster to learn
and require less tuning. This may be a significant advantage since
the ultimate goal is to make knockoffs easily accessible to
researchers from different fields. A computationally lighter
alternative is proposed in \cite{liu2018auto}, which relies on the
variational autoencoder \cite{kingma2013auto} to generate knockoff
copies. Since our work was developed in parallel\footnote{The results of this paper were first discussed at the University of California, Los Angeles, during the Green Family Lectures on September 27, 2018.} to that of
\cite{anonymous2019knockoffgan, liu2018auto}, we are not including
these recent proposals in our simulation studies.  Instead, we will
compare our method to well-established alternatives.

\section{Model-X knockoffs} \label{sec:knockoffs}
Since the scope of our work depends on properties of model-X
knockoffs, we begin by rehearsing some of the key features of the
existing theory.  For any $X \in \RR^p$ sampled from a distribution
$P_X$, the random vector $\tilde{X} \in \RR^p$ is said to be a
knockoff copy of $X$ \cite{candes2016panning} if the joint law of
$(X, \tilde{X})$ obeys 
\begin{align} \label{eq:knock-2} (X, \tilde{X}) \overset{d}{=} (X,
  \tilde{X})_{\text{swap}(j)} \quad \text{for each } j \in
  \{1,\ldots,p\};
\end{align}
here, the symbol $\overset{d}{=}$ indicates equality in distribution
and $(\cdot)_{\text{swap}(j)}$ is defined as the operator swapping
$X_j$ with $\tilde{X}_j$; if $p = 3$ and $j = 2$,
$(X_1,X_2,X_3,\tilde{X}_1, \tilde{X}_2, \tilde{X}_3)_{\text{swap}(j)}
= (X_1,\tilde{X}_2,X_3,\tilde{X}_1, {X}_2, \tilde{X}_3)$. Knockoffs
play a key role in controlled variable selection, by serving as
negative controls that allow one to estimate and limit the number of
false positives in the variable selection problem defined below.

Consider $n$ observations $\{X^i, Y^i\}_{i=1}^n$, with each $X^i = (X^i_1, \ldots, X^i_p) \in \RR^p$ assumed to be drawn independently from a known $P_X$, and the associated label $Y^i \in \RR$ drawn from an unknown conditional distribution $P_{Y \mid X}$. The goal is to identify a subset of important components of $X$ that affect $Y$. In order to state this objective more precisely, one refers to $X_j$ as unimportant if
\begin{align*}
  Y \indep X_j \mid X_{-j},
\end{align*}
where $X_{-j}$ indicates the remaining $p-1$ variables after $X_j$ is
excluded. The true null hypotheses $\mathcal{H}_0$ is the set of all
variables that are unimportant; in words, $X_j$ is not important if it
is conditionally independent of the response $Y$ once we know the
value of $X_{-j}$.  Put differently, $X_j$ is not important if it does
not provide any additional information about $Y$ beyond what is
already known.  While searching for a subset $\hat{\mathcal{S}}$ that
includes the largest possible number of important variables in
$\mathcal{H}_1 = \{1,\ldots,p\} \setminus \mathcal{H}_0$, one wishes
to ensure that the false discovery rate,
\begin{align*}
  \text{FDR} 
  & = \EE\left[ \frac{ |\hat{\mathcal{S}} \cap \mathcal{H}_0|  }{ |\hat{\mathcal{S}} \lor 1| } \right],
\end{align*}
remains below a nominal level $q \in (0,1)$, e.g.~$q=0.1$. The false discovery rate is thus defined as the expected fraction of selected variables that are false positives. 

The approach of \cite{candes2016panning} provably controls the false discovery rate without placing any restrictions on the conditional likelihood of $Y \mid X$, which can be arbitrary and completely unspecified. The first step in their method consists of generating a knockoff copy $\tilde{X}$ for each available sample of $X$, before looking at $Y$, such that both \eqref{eq:knock-2} is satisfied and 
$ Y \indep \tilde{X} \mid X$.
Some measures of feature importance $Z_j$ and $\tilde{Z}_j$ are then
evaluated for each $X_j$ and $\tilde{X}_j$, respectively. For this
purpose, almost any available method from statistics and machine
learning can be applied to the vector of labels $\Y$ and the augmented
data matrix $[\X, \tilde{\X}] \in \RR^{n \times 2p}$, with the only
fundamental rule that the original variables and the knockoffs should
be treated equally; this is saying that the method should not use any
information revealing which variable is a knockoff and which is not.
Examples include sparse generalized linear models
\cite{candes2016panning, sesia2017gene}, random forests
\cite{Gao2018}, support vector machines and deep neural networks
\cite{gimenez2018knockoffs, lu2018deeppink}. Each pair of $Z_j$ and
$\tilde{Z}_j$ is then combined through an antisymmetric function into
the statistics $W_j$, e.g.~$W_j = Z_j-\tilde{Z}_j$. By construction, a
large and positive value of $W_j$ suggests evidence against the $j$th
null hypothesis, while unimportant variables are equally likely to be
positive or negative. Under this choice of $W_j$, it can be shown that
exact control of the false discovery rate below the nominal level $q$
can be obtained by selecting
$\hat{\mathcal{S}} = \left\{ j : W_j \geq \tau_q \right\}$, where
\begin{align*}
  \tau_q = \min \left\{ t > 0 : \frac{1 + |\{ j : W_j \leq -t\} | }{|\{ j : W_j \geq t\}|} \leq q \right\}.
\end{align*}
The numerator in the expression above can be understood as a
conservative estimate of the number of false positives above the fixed
level $t$. This adaptive significance threshold is that
first proposed in the knockoff filter of \cite{barber2015}, while the choice of the test statistics $W_j$ may be different \cite{candes2016panning}. 

The validity of the false discovery rate control relies entirely on the exact knowledge of $P_X$ and our ability to generate $\tilde{X}$ satisfying \eqref{eq:knock-2}. Even though procedures that can sample exact knockoff copies have been previously derived for a few special classes of $P_X$ such as multivariate Gaussian distributions \cite{candes2016panning} and hidden Markov models \cite{sesia2017gene}, the general case remains algorithmically challenging. 
This difficulty arises because \eqref{eq:knock-2} is much more stringent than a first look may suggest. For instance, obtaining new independent samples from $P_X$ or permuting the rows of the data matrix would only ensure that $(X_1, X_2)$ is equal in distribution to $(\tilde{X}_1, \tilde{X}_2)$, while the analogous result would not hold between $(X_1, X_2)$ and $(X_1, \tilde{X}_2)$. At the same time, the latter property is crucial since a null variable and its knockoff must be able to explain on average the same fraction of the variance in the response. 
The practical approximate solution described in \cite{candes2016panning} consists of relaxing the condition in \eqref{eq:knock-2} as to match only the first two moments of the distributions on either side.
In this weaker sense, $\tilde{X}$ is thus said to be a {\em second-order knockoff} copy of $X$ if the two random vectors have the same expected value and their joint covariance matrix is equal to
\begin{align} \label{eq:second-order}
  \text{Cov}\left[(X,\tilde{X})\right]
  & = \begin{bmatrix}
    \Sigma & \Sigma - \text{diag}(s) \\
    \Sigma - \text{diag}(s) & \Sigma
  \end{bmatrix},
\end{align}
where $\Sigma$ is the covariance matrix of $X$ under $P_X$ and $s$ is
any $p$-dimensional vector selected in such a way that the matrix in
the right-hand side is positive semidefinite. The role of $s$ is to
make $\tilde{X}$ as uncorrelated with $X$ as possible, in order to
increase statistical power during variable selection. Therefore, the
value of $s$ is typically chosen to be as large as possible 
\cite{candes2016panning}.  The weaker form of exchangeability in
\eqref{eq:second-order} is reminiscent of the notion of fixed-design
knockoffs from \cite{barber2015} and it can be practically implemented
by approximating the distribution of $X$ as multivariate Gaussian
\cite{candes2016panning}. This approximation often works well in
practice, even though it is in principle insufficient to guarantee
control of the false discovery rate under the general conditions of
the model-X framework \cite{barber2018robust}.  In this paper, we
build upon the work of \cite{candes2016panning} and
\cite{barber2018robust} to obtain higher-order knockoffs that can
achieve a better approximation of \eqref{eq:knock-2} using modern
techniques from the field of deep generative models.


\section{Deep generative models} \label{sec:deep-models}
Replicating the underlying distribution of a data source is an essential task of statistical machine learning that can be broadly described as follows. Given $n$ independent $p$-dimensional samples $\{X^i\}_{i=1}^n$ from an unknown distribution $P_X$, a generative model approximating the true $P_X$ is sought in order to synthesize new observations that could plausibly belong to the training set, while being sufficiently different to be non-trivial. Several well known techniques have been developed to tackle this problem, some of which are based on hidden Markov models \cite{baum1966statistical}, Gaussian mixture models \cite{nasrabadi2007pattern} or Boltzmann machines \cite{ackley1985learning}. In recent years, such traditional approaches have been largely replaced by neural networks, with two popular examples being variational autoencoders \cite{kingma2013auto, burda2015importance, sonderby2016ladder, tolstikhin2017wasserstein} and generative adversarial networks \cite{goodfellow2014generative, makhzani2015adversarial, nowozin2016f, chen2016infogan, mescheder2017adversarial, arjovsky2017wasserstein, karras2017progressive}. These are based on a parametric non-linear function $f_\theta(V)$ that maps an input noise vector $V$ to the sample domain of $X$. The parameters in $\theta$ represent the collection of weights and biases defining the neural network and they need to be learned from the available data. The function thus defined is deterministic for any fixed realization of the noise and, with an appropriate choice of $\theta$, it propagates and transforms the noise in $V$ to obtain a random variable $f_\theta(V)$ approximately distributed as $X$. 

Training deep generative models is computationally difficult, and considerable  effort has been dedicated to the development of practical algorithms that can find good solutions. For instance, the popular variational method, which lies at the heart of the autoencoder in \cite{kingma2013auto}, proceeds by maximizing a traceable lower bound on the log-likelihood of the training data. In contrast, generative adversarial networks strive to minimize the inconsistencies of the generated samples with the original ones, by formulating the learning task as a two-player game \cite{goodfellow2014generative}. As the generator $f_\theta(V)$ attempts to produce realistic samples, an antagonistic discriminator tries to recognize them. Since the discriminator is defined as a deep binary classifier with a differentiable loss function, the two networks can be simultaneously trained by gradient descent, until no further gain can be made on either side. Even though generative adversarial networks have enjoyed a great deal of success \cite{goodfellow2014generative, makhzani2015adversarial, nowozin2016f, chen2016infogan, mescheder2017adversarial, arjovsky2017wasserstein, karras2017progressive}, non-convex minimax optimization is notoriously complex \cite{arjovsky2017towards}. More recent alternatives mitigate the issue by replacing the classifier with a simpler measure of the distance between the distributions of the original and the simulated samples \cite{li2015generative, dziugaite2015training, li2017mmd, srivastava2018ratio, arbel2018gradient, binkowski2018demystifying}. The remaining part of this section is dedicated to reviewing the basics of some of the latter approaches, upon which we will begin to develop a knockoff machine. 

The discriminator component of a deep generative model faces the
following challenge. Given two sets of independent observations
$\{X^i\}_{i=1}^n$ and $\{Z^i\}_{i=1}^n$, respectively drawn from some
unknown distributions $P_X$ and $P_Z$, it must be verified whether
$P_X = P_Z$. This multivariate two-sample problem has a long history
in the statistics literature and many non-parametric tests have been
proposed to address it \cite{bickel1969distribution,
  friedman1979multivariate, schilling1986multivariate,
  henze1988multivariate, friedman2004multivariate, gretton2012kernel,
  szekely2013energy}. In particular, the work of
\cite{gretton2012kernel} introduced a test statistic, called the
maximum mean discrepancy, whose desirable computational properties
have inspired the development of generative moment matching networks
\cite{li2015generative, dziugaite2015training}. The relevant key idea
is to quantify the discrepancy between the two distributions in terms
of the largest difference in expectation between $\phi(X)$ and
$\phi(Z)$, over functions $\phi$ mapping the random variables into the
unit ball of a reproducing kernel Hilbert space
\cite{gretton2012kernel}. Fortunately, this abstract characterization
can be made explicit with the kernel trick \cite{gretton2012kernel},
leading to the practical utilization described below. 

Let $X, X', Z, Z'$ be independent samples drawn from $P_X$ and
 $P_Z$, respectively, and define the maximum mean discrepancy between $P_X$ and $P_Z$ as
 \begin{align} \label{eq:def-MMD}
   \mathcal{D}_{ \text{MMD}}(P_X, P_Z) 
   & = \EE_{X,X'} \left[k(X,X')\right] - 2\EE_{X,Z} \left[k(X,Z)\right] + \EE_{Z,Z'} \left[k(Z,Z')\right],
 \end{align}
 where $k$ is a kernel function.
 If the characteristic kernel of a reproducing kernel
 Hilbert space \cite{gretton2012kernel} is used, it can be shown that the quantity
 in \eqref{eq:def-MMD} is equal to zero if and only if $P_X=P_Z$. 
 Concretely, valid choices of $k$ include the
 common Gaussian kernel, $k(X,X') = \exp\{-\|X - X'\|_2^2/(2\xi^2)\}$,
 with bandwidth parameter $\xi>0$, and mixtures of such. Furthermore,
 the maximum mean discrepancy is always non-negative and it can be
 estimated from finite samples $\X,\Z \in \RR^{n \times p}$ in an
 unbiased fashion via 
 \begin{align} \label{eq:MMD-hat}
   \widehat{\mathcal{D}}_{ \text{MMD}}(\X, \Z)
   & = \frac{1}{n(n-1)} \sum_{i=1}^n \sum_{j \neq i} k(X^i,X^j) -
     \frac{2}{n^2} \sum_{i=1}^n \sum_{j=1}^n k(X^i, Z^j) +
     \frac{1}{n(n-1)}\sum_{i=1}^n\sum_{j \neq i} k(Z^i,Z^j), 
 \end{align}
see \cite{gretton2012kernel}. 
Since the expression in \eqref{eq:MMD-hat} is easily computable and differentiable, it can serve as the objective function of a deep generative model, effectively replacing the discriminator required by generative adversarial networks \cite{li2015generative, dziugaite2015training}. The generator is then trained on $\X$ to produce samples $\Z$ that minimize \eqref{eq:MMD-hat}, by applying the standard techniques of gradient descent. This idea can also be repurposed to develop a knockoff machine, as discussed in the next section.


\section{Deep knockoff machines} \label{sec:machines}
\subsection{Overview} \label{sec:machines-general}

A knockoff machine is defined as a random mapping
$f_\theta$ that takes as input a random $X \in \reals^p$, an independent noise vector
$V \sim \mathcal{N}(0,I) \in \RR^p$ and returns an approximate 
knockoff copy $\tilde{X} = f_{\theta}(X,V) \in \reals^p$. 
The machine is characterized by a set of parameters $\theta$
and it should be designed such that the joint distribution of 
$(X,\tilde{X})$ deviates from \eqref{eq:knock-2} as little as possible.
If the original variables follow a multivariate
Gaussian distribution, i.e.~$X \sim \mathcal{N}(0,\Sigma)$, 
a family of machines generating exact knockoffs is given by
\begin{align} \label{eq:machine-gaussian}
  f_\theta(X,V) 
  & = X - X\Sigma^{-1}\text{diag}\{s\} + \left(2\text{diag}\{s\} - \text{diag}\{s\}\Sigma^{-1}\text{diag}\{s\}\right)^{1/2}V,
\end{align}
for any choice of the vector $s$ that keeps the matrix multiplying $V$
positive-definite \cite{candes2016panning}. In practice, the value of
$s$ is typically determined by solving a semi-definite program
\cite{candes2016panning}, see Section~\ref{sec:machine-details}. By contrast, the algorithm for sampling knockoff copies of hidden Markov models in \cite{sesia2017gene} cannot be easily represented as an explicit function $f_\theta$. This difficulty should be expected for various other choices of $P_X$, and an analytic derivation of $f_{\theta}$ seems intractable in general.  

\begin{figure}[!htb]
  \centering
    \includegraphics[width=0.42\textwidth]{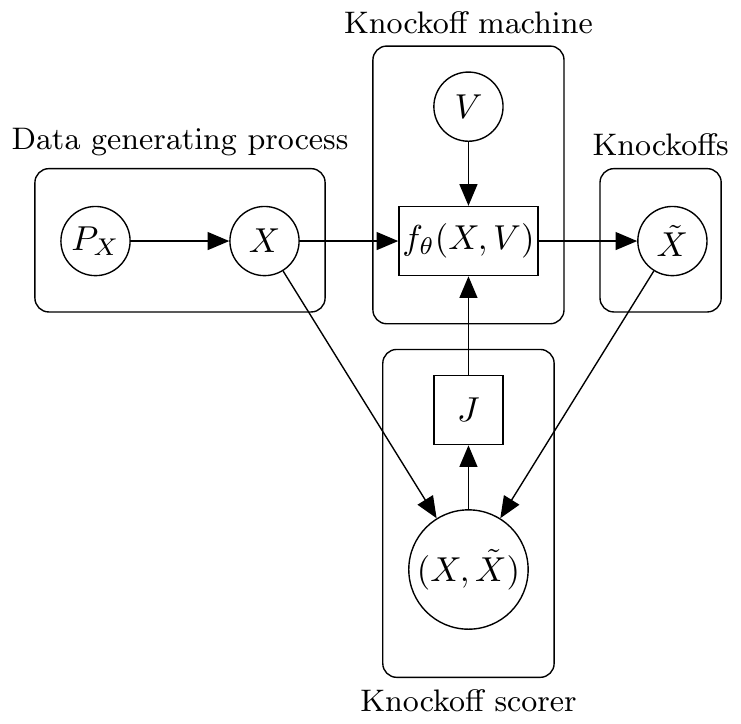}
    \caption{Schematic
      representation of the learning mechanism of a knockoff
      machine. The arrows indicate the flow of information between the source of data, 
      the machine and the knockoff scoring function.}
  \label{fig:machine}
\end{figure}

In order to develop a flexible machine that can sample knockoffs for
arbitrary and unknown distributions $P_X$, we assume $f_\theta$
to take the form of a deep neural network, as described in
Section~\ref{sec:machine-details}. The values of its parameters will
be estimated on the available observations of $X$ by solving a
stochastic optimization problem. An overview of our approach is
visually presented in Figure~\ref{fig:machine} and it can be
summarized as follows: the machine is provided with $n$ realizations of the random vector $X$, independently sampled from an unknown underlying distribution $P_X$. For any fixed configuration of $\theta$, each $\tilde{X}^i$ is computed as a function of the corresponding input $X^i$ and the noise $V^i$, for $i \in \{1,\ldots,n\}$. The latter ($V^i$) is independently resampled for each observation and each time the machine is called. A scoring function $J$ examines the empirical distribution of $(X,\tilde{X})$ and quantifies its compliance with the exchangeability property in \eqref{eq:knock-2}. After each such iteration, the parameters $\theta$ are updated in the attempt to improve future scores. Ideally, upon successful completion of this process, the machine should be ready to generate approximate knockoff copies $\tilde{X}$ for new observations of $X$ drawn from the same $P_X$. A specific scoring function that can generally lead to high-quality knockoffs will be defined below.

\subsection{Second-order machines} \label{sec:machines-second-order}

We begin by describing the training of a special knockoff machine
that is interesting for expository
purposes. Suppose that instead of requiring the joint distribution of $(X,\tilde{X})$ to satisfy \eqref{eq:knock-2}, we would be satisfied with obtaining second-order knockoffs.
In order to incentivize the machine to produce $\tilde{X}$ such that $\mathbb{E}[X]=\mathbb{E}[\tilde{X}]$ and the joint covariance matrix of $(X,\tilde{X})$ satisfies \eqref{eq:second-order}, we consider a simple loss function that computes a differentiable measure of its compatibility with these requirements.
For the sake of notation, we let $\hat{G}$ indicate the empirical covariance matrix of $(X,\tilde{X}) \in \RR^{2p}$, which takes the following block form:
\begin{align} \label{eq:covariance-joint}
\hat{G}
& = \begin{bmatrix}
\hat{G}_{XX} & \hat{G}_{X\tilde{X}} \\
\hat{G}_{X\tilde{X}} & \hat{G}_{\tilde{X}\tilde{X}}
\end{bmatrix}.
\end{align}
Above, $\hat{G}_{XX}, \hat{G}_{\tilde{X}\tilde{X}} \in \reals^{p \times p}$ are the empirical covariance matrices of $X,\tilde{X}$, respectively.
Then we define
\begin{align} \label{eq:J-second-order}
J_{ \text{second-order}}(\X, \tilde{\X}) 
& = \lambda_1 \frac{\| \frac{1}{n} \sum_{i=1}^{n} (X^i - \tilde{X}^i) \|_2^2}{p} + 
  \lambda_2 \frac{\| \hat{G}_{XX} - \hat{G}_{\tilde{X}\tilde{X}}\|_F^2}{\| \hat{G}_{XX}\|_F^2} + 
  \lambda_3 \frac{\| M \circ (\hat{G}_{XX} - \hat{G}_{X\tilde{X}}) \|_F^2}{\| \hat{G}_{XX}\|_F^2}.
\end{align} 
Here, the symbol $\circ$ indicates element-wise multiplication, while $M = E - I \in \RR^{p \times p}$, with $E$ being a matrix of ones and $I$ the identity matrix. For simplicity, the weights $\lambda_1, \lambda_2, \lambda_3$ will be set equal to one throughout this paper. The first term in \eqref{eq:J-second-order} penalizes differences in expectation, while the second and third terms encourage the matching of the second moments. Smaller values of this loss function intuitively suggest that $\tilde{X}$ is a better second-order approximate knockoff copy of $X$.
Since $J$ is smooth, a second-order knockoff machine can be trained with standard techniques of stochastic gradient descent.

As we mentioned earlier, knockoffs are not uniquely defined, and it is desirable to make $\tilde{X}$ as different as possible from $X$. There are various ways of encouraging a machine to seek this outcome, and a practical solution inspired by \cite{candes2016panning} consists of adding a regularization term to the loss function, penalizing large pairwise empirical correlations between $X$ and $\tilde{X}$:
\begin{align} \label{eq:J-power}
J_{ \text{decorrelation}}(\X, \tilde{\X}) 
& = \sum_{j=1}^{p} \widehat{\text{corr}}(X_j, \tilde{X}_j).
\end{align}
Each term  above is defined as the empirical estimate of the Pearson correlation coefficient for the $j$th columns of $\X$ and $\tilde{\X}$. In summary, this describes a new general procedure for sampling approximate second-order knockoffs. Compared to the original method in \cite{candes2016panning}, the additional computational burden of fitting a neural network is significant. However, the tools developed in this section are valuable because they can be generalized beyond the second-order setting, as discussed next.

\subsection{Higher-order machines} \label{sec:machines-higher-order}

In order to build a general knockoff machine, one must be able to
precisely quantify and control the deviation from exchangeability: the difference in distribution between $(X,\tilde{X})$ and $(X,\tilde{X})_{\text{swap}(j)}$ for each 
$j \in \{1,\ldots,p\}$.  For this purpose, we deploy the
maximum mean discrepancy metric from
Section~\ref{sec:deep-models}. In order to obtain an unbiased estimate, 
we randomly split the data into a partition $\X',\X'' \in \reals^{n/2 \times p}$
and define the corresponding output of the machine as $\tilde{\X}',\tilde{\X}''$. Then,
it is natural to seek a machine that targets 
\begin{align*}
  \sum_{j=1}^{p} \widehat{\mathcal{D}}_{ \text{MMD}} \left[ (\X',\tilde{\X}'), (\X'',\tilde{\X}'')_{\text{swap}(j)} \right].
\end{align*}
Above, $\widehat{\mathcal{D}}_{ \text{MMD}}$ stands for the empirical estimate in \eqref{eq:MMD-hat} of the maximum mean discrepancy, evaluated with a Gaussian kernel. 
Intuitively, the above quantity is minimized in expectation if the knockoffs are exchangeable according to \eqref{eq:knock-2}. This idea will be made more precise below, for a slightly modified objective function.
We refer to this solution as a higher-order knockoff machine because the expansion of the Gaussian kernel into a power series leads to a characterization of \eqref{eq:def-MMD} in terms of the distance between vectors containing all higher-moments of the two distributions \cite{cotter2011explicit, gretton2012kernel}. Therefore, our approach can be interpreted as a natural generalization of the method in \cite{candes2016panning}. 

Since computing $\widehat{\mathcal{D}}_{ \text{MMD}}$ at each
iteration may be expensive (there are $p$ swaps), in practice we will only consider two swaps and
ask the machine to minimize
\begin{align} \label{eq:J-MMD}
  J_{ \text{MMD}}(\X, \tilde{\X})
  & = \widehat{\mathcal{D}}_{ \text{MMD}} \left[ (\X',\tilde{\X}'), (\tilde{\X}'', \X'')\right] + 
    \widehat{\mathcal{D}}_{ \text{MMD}} \left[ (\X',\tilde{\X}'), (\X'',\tilde{\X}'')_{\text{swap}(S)} \right],
\end{align}
where $S$ indicates a uniformly chosen random subset of $\{1,\ldots,p\}$ such that $j \in S$ with probability $1/2$.
The following result confirms that the objective function in  \eqref{eq:J-MMD} provides a sensible guideline for training knockoff machines.

\begin{theorem} \label{thm:machine-loss}
Let $\X \in \reals^{n \times p}$ be a collection of independent observations drawn from $P_X$, and define $\tilde \X$ as the corresponding random output of a fixed machine $f_\theta$. Then for $J_{ \text{\em MMD}}$ defined  as in \eqref{eq:J-MMD}, 
$$\E{J_{ \text{\em MMD}}(\X, \tilde{\X}) } \geq 0.$$
Moreover, equality holds if and only if the machine produces valid knockoffs for $P_X$. Above, the expectation is taken over $\X$, the noise in the knockoff machine, and the random swaps in the loss function.
\end{theorem}

With a finite number of observations available, stochastic gradient descent aims to minimize the  expectation of \eqref{eq:J-MMD} conditional on the data. This involves solving a high-dimensional non-convex optimization problem that is difficult to study theoretically. Nonetheless, effective algorithms exist and a weak form of convergence of stochastic gradient descent is established in Section~\ref{sec:theory}. Therefore, these results provide a solid basis for our proposed method.  

The full objective function of a knockoff machine may also include the quantities from \eqref{eq:J-second-order} and \eqref{eq:J-power}, as a form of regularization, thus reading as
 \begin{align} \label{eq:J-loss}
   J(\X, \tilde{\X})
   & = \gamma J_{ \text{MMD}}(\X, \tilde{\X}) + \lambda J_{\text{second-order}}(\X, \tilde{\X}) + \delta J_{ \text{decorrelation}}(\X, \tilde{\X}).
 \end{align}
 In the special case of $\gamma=0$, a second-order machine is recovered, while $\delta=0$ may lead to knockoffs with little power. The second-order penalty may appear redundant because $J_{ \text{MMD}}$ already penalizes discrepancies in the covariance matrix, as well as in all other moments. However, we have observed that setting $\lambda > 0$ often helps to decrease the amount of time required to train the machine.
For optimal performance, the hyperparameters should be tuned to the specific data distribution at hand. For this purpose, we discuss practical tools to measure goodness of fit later in Section~\ref{sec:tuning}. Meanwhile, for any fixed choice of $(\gamma,\lambda,\delta)$, the learning strategy is summarized in Algorithm~\ref{alg:machine-training}. 

\SetKwInput{KwInput}{Input}
\SetKwInput{KwOutput}{Procedure}
\SetKwInput{KwResult}{Output}

\begin{algorithm}[h!]
	\SetAlgoLined
	\KwInput{$\X\in\reals^{n\times p}$ -- Training data.}
	\myinput{$\gamma$ -- Higher-order penalty hyperparameter.}
	\myinput{$\lambda$ -- Second-order penalty hyperparameter.}
	\myinput{$\delta$ -- Decorrelation penalty hyperparameter.}        
	\myinput{$\theta_1$ -- Initialization values for the weights and biases of the network.}
	\myinput{$\mu$ -- Learning rate.}
	\myinput{$T$ -- Number of iterations.}
		
	\KwResult{$f_{\theta_T}$ -- A knockoff machine.}
	\vspace{5pt}
	\KwOutput{}
	\For{$ t = 1:T $}{
		\vspace{5pt}
                Sample the noise realizations:
                $V^i\sim \mathcal{N}(0,I)$, for all $1\leq i \leq n$\;
		\vspace{5pt}
		Randomly divide $\X$ into two disjoint mini-batches $\X', \X''$\;	
		\vspace{5pt}
                Pick a subset of swapping indices $S \subset \{1,\ldots,p\}$ uniformly at random\;
		\vspace{5pt}
		Generate the knockoffs as a deterministic function of $\theta$: \\
		$\tilde{X}^i = f_{\theta_t}(X^i,V^i)$, for all $1\leq i \leq n$\;
		\vspace{5pt}
                Evaluate the objective function, using the batches and swapping indices fixed above: \\
		$J_{\theta_t}(\X, \tilde{\X}) = \gamma J_{ \text{MMD}}(\X, \tilde{\X}) + \lambda J_{\text{second-order}}(\X, \tilde{\X}) + \delta J_{ \text{decorrelation}}(\X, \tilde{\X})$\;
		\vspace{5pt}
                Compute the gradient of $J_{\theta_t}(\X, \tilde{\X})$, which is now a deterministic function of $\theta$\;
		\vspace{5pt}
		Update the parameters: 
		$\theta_{t+1} = \theta_{t} - \mu \nabla_{\theta_t}J_{\theta_t}(\X, \tilde{\X})$\;
	}
	\caption{Training a deep knockoff machine}
	\label{alg:machine-training}
\end{algorithm}

Alternative types of knockoff machines could be based on different
choices of kernel or other measures of the discrepancy between two
distributions. An intuitive option would be the Kullback-Leibler
divergence \cite{jiang2018approximate}, which appears at first sight
to be a natural choice. In fact, a connection has been shown in
\cite{barber2018robust} between this and the worst-case inflation of
the false discovery rate that may occur if the variable selection
relies on inexact knockoffs. In recent years, some empirical
estimators of this divergence have been proposed in the literature on
deep generative models
\cite{nguyen2010estimating,jiang2018approximate}, which could also be
employed for our purposes.

\subsection{Analysis of the optimization algorithm} \label{sec:theory}

In this section, we study the behavior of Algorithm
\ref{alg:machine-training}, establishing a weak form of
convergence. For simplicity, we focus on the machine
defined by the loss function in \eqref{eq:J-loss} with
$(\gamma,\lambda,\delta)=(1,0,0)$. The other cases can
be treated similarly and they are omitted in the interest of space.
In order to facilitate the
exposition of our analysis, we begin by introducing some helpful
notations. Let $\X'_t$ and $\X''_t$ denote a randomly chosen 
partition of the fixed training set $\X \in \reals^{n \times p}$.
 The state of the learning algorithm at time $t$ is
fully described by $\zeta_t = (\X'_t, \X''_t, \theta_t)$, where
$\theta_t$ is the current configuration of the machine
parameters. Conditional on the noise realizations
$\varepsilon_t = (\V_t', \V_t'')$ and the randomly chosen set of
swapping indices $S_t$, the objective
\begin{align*} 
  J_{ \text{MMD}}(\X'_t, \tilde{\X}'_t, \X''_t, \tilde{\X}''_t, S_t),
\end{align*}
is a deterministic
function of $\theta_t$
since $\tilde \X'_t = f_{\theta_t}(\X'_t, \V'_t)$ and
$\tilde \X''_t = f_{\theta_t}(\X''_t, \V''_t)$. Above,
$J_{ \text{MMD}}$ is written with a
slight, but clarifying, abuse of the notation in \eqref{eq:J-MMD}.  At
this point, we can also define
\begin{align} \label{eq:expected-loss}
  J_{\theta_t} = \Ec{J_{ \text{MMD}}(\X'_t, \tilde{\X}'_t, \X''_t, \tilde{\X}''_t,S_t)}{\zeta_t},
\end{align}
with the expectation taken over the noise $\varepsilon_t$ and the choice of $S_t$. Let us also define $\nabla J_{\theta_t}$ as the gradient of \eqref{eq:expected-loss} with respect to $\theta_t$. In practice, this quantity is approximated by sampling one realization of $\varepsilon_t$ and a set of swapping indices $S_t$, then computing the following unbiased estimate:
\begin{align} \label{eq:gt}
  g_t = \nabla J_{ \text{MMD}}(\X'_t, f_{\theta_t}(\X'_t, \V'_t), \X''_t, f_{\theta_t}(\X''_t, \V''_t), S_t).
\end{align}
Since the function is deterministic because all random variables in \eqref{eq:gt} have been observed by the algorithm, backpropagation can be used to calculate the gradient on the right-hand-side.
This gradient is then used to update the machine parameters in the next step, through $\theta_{t+1} = \theta_t - \mu g_t$, where $\mu$ is the learning rate. Under standard regularity conditions, we can follow the strategy of \cite{sanjabi2018solving} to show that the algorithm tends to approach a stationary regime. In particular, we assume the existence of a finite Lipschitz constant $L$ such that, for all $\theta', \theta''$ and all possible values of the data batches $\X', \X''$, 
\begin{align*}
  \left\| \nabla \Ec{J_{ \text{MMD}}(\X', \tilde{\X}', \X'', \tilde{\X}'', S)}{\X', \X'', \theta'} - \nabla \Ec{J_{ \text{MMD}}(\X', \tilde{\X}', \X'', \tilde{\X}'', S)}{\X', \X'', \theta''} \right\|_2
  & \leq L \|\theta' - \theta''\|_2,
\end{align*}
and we define
\begin{align*}
\Delta = \frac{2}{L} \sup \left(J_{\theta_1} - J^*\right).
\end{align*}
Above, $J_{\theta_1}$ indicates the expected loss \eqref{eq:expected-loss} at the first step, conditional on the data and the initialization of $\theta$. The supremum is taken over all possible values of the data and the initial $\theta$.
The value of $J^*$ is defined as a uniform lower bound on $J_{\theta_t}$.   Following the result of \cite{gretton2012kernel} that bounds the empirical estimate of the maximum mean discrepancy from below, 
\begin{align*}
  & \widehat{\mathcal{D}}_{ \text{MMD}}(\X, \Z)
    \geq - \frac{1}{n(n-1)} \sum_{i=1}^{n} \left[ k(X^i,X^i) + k(Z^i,Z^i) - k(X^i,Z^i)\right],
  & \forall  \X, \Z \in \reals^{n \times p}, 
\end{align*}
we can conclude that a finite value of $J^*$ can be determined from the data.

\begin{theorem} \label{thm:machine-convergence} Consider a fixed
  training set $\X \in \reals^{n \times p}$ and adopt the notation
  above.  Assume that the gradient estimates have uniformly bounded
  variance; that is, 
\begin{align*}
\Ec{ \| g_t - \nabla J_{\theta_t} \|_2^2 }{ \zeta_t} \leq \sigma^2, \qquad \forall t \leq T,
\end{align*}
for some $\sigma^2 \in \reals$. Then for any initial state $\zeta_1$ of the machine and a suitable value of the constant $\Delta$ defined above,
\begin{align*}
\frac{1}{T}\sum_{t=1}^T\Ec{ \|\nabla J_{\theta_t}\|_2^2}{\zeta_1} & \leq \frac{1}{T} \frac{L\Delta}{\mu \left( 2 - L\mu\right)} + \frac{L\sigma^2 \mu}{\left( 2 - L\mu\right)}.
\end{align*}
In particular, choosing $\mu = \min\left\{\frac{1}{L},  \frac{\mu_0}{\sigma  \sqrt{T} } \right\}$ for some $\mu_0>0$ gives
\begin{align*}
\frac{1}{T}\sum_{t=1}^T\Ec{ \|\nabla J_{\theta_t}\|_2^2}{\zeta_1} & \leq \frac{L^2\Delta}{T} + \left(\mu_0 + \frac{\Delta}{\mu_0}\right)\frac{L\sigma}{\sqrt{T}}.
\end{align*}
\end{theorem}

In a nutshell, Theorem \ref{thm:machine-convergence} states that the squared norm of the gradient of the loss function \eqref{eq:expected-loss} decreases on average as $\mathcal{O}(T^{-1/2})$ when $T \to \infty$. This can be interpreted as a weak form of convergence that is not necessarily implying that $\theta_t$ will reach a fixed point. One could also follow the strategy of \cite{ghadimi2013stochastic} instead of \cite{sanjabi2018solving} to obtain a closely related result, guaranteeing that the norm of the gradient will be small at a sufficiently large and randomly chosen stopping time. It would of course be more desirable to establish the convergence in a stronger sense, perhaps to a local minimum; however, this is difficult and we are not aware of any similar results in the literature on deep moment matching networks. It should be noted that our assumption that the gradient estimates have uniformly bounded variance is not as strong as requiring the gradients to be uniformly bounded. The work of \cite{ithapu2017architectural} provides explicit bounds in several special instances of single and multi-layer neural networks. However, we choose not to validate this assumption in our knockoff machines for two reasons. First, it is standard in the literature \cite{ghadimi2013stochastic, sanjabi2018solving}; second, a proof would need to rely heavily on a specific architecture and loss function. In practice, we observed that a learning rate in the typical range between $0.001$ and $0.01$ works well.

\subsection{Implementation details} \label{sec:machine-details}

\begin{figure}[!htb]
  \centering
    \includegraphics[width=0.73\textwidth]{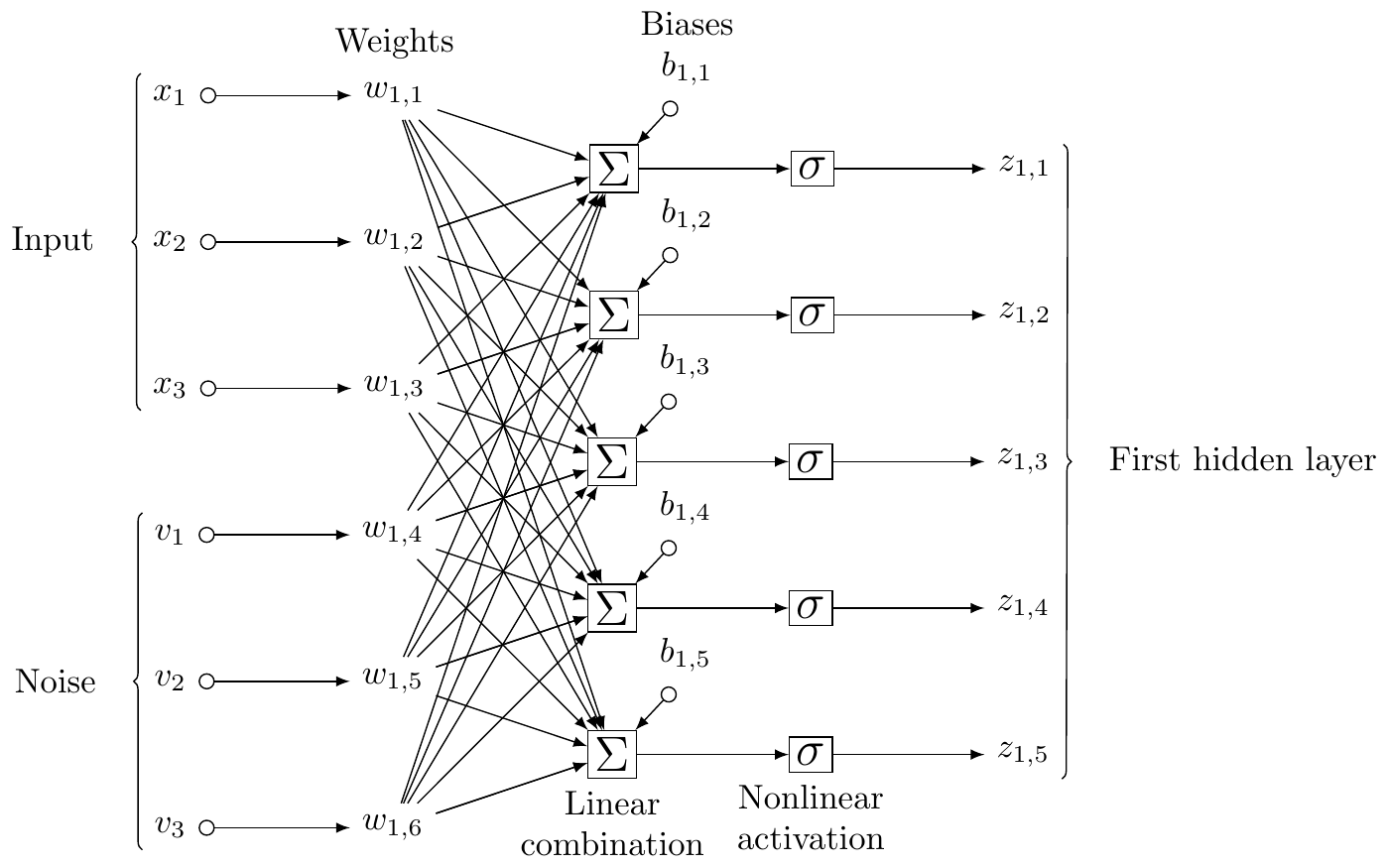}
  \caption{Connections in the input layer of a knockoff machine. This layer takes as input 3 input variables and 3 noise instances, producing 5 latent variables.}\label{fig:network-input}
\end{figure}

\begin{figure}[!htb]
    \centering
    \begin{subfigure}[t]{0.4255\textwidth}
        \centering
        \includegraphics[width=\textwidth]{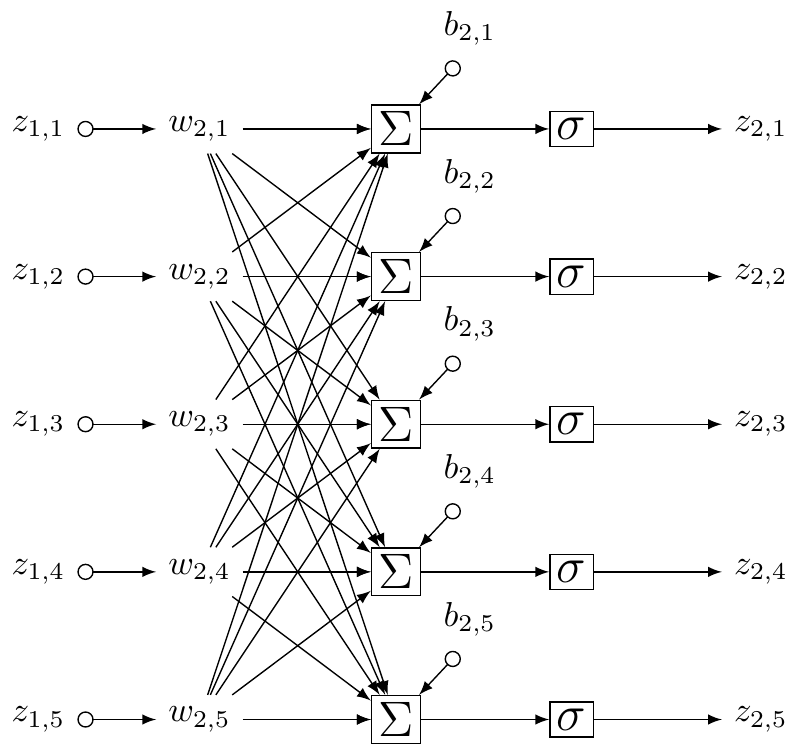}
        \caption{}  \label{fig:network-hidden}
    \end{subfigure}
    ~
    \begin{subfigure}[t]{0.5\textwidth}
        \centering
        \includegraphics[width=\textwidth]{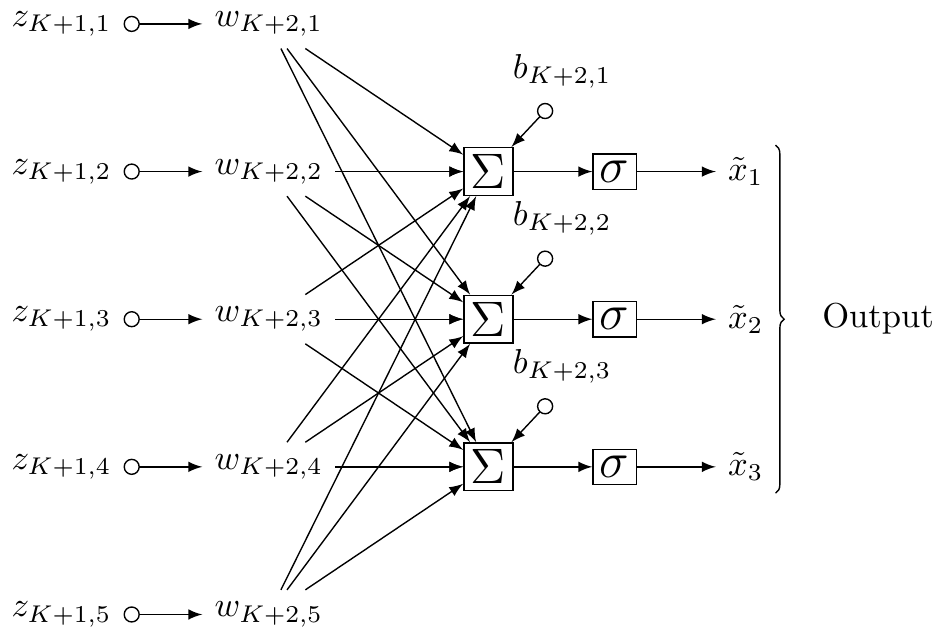}
        \caption{}  \label{fig:network-output}
    \end{subfigure}
  \caption{Visualization of the connections in the hidden layers (a) and in the output layer (b) of the knockoff machine from Figure~\ref{fig:network-input}. The complete machine encodes a knockoff generating function $f_{\theta}(X,V)$ by concatenating an input layer, $K$ hidden layers and an output layer.}
\end{figure}

The construction of deep knockoff machines allows considerable freedom in the precise form of $f_\theta$. In general, neural networks can be implemented following a multitude of different architectures, and the final choice is often guided by the experience of the practitioners. For the purpose of this paper, we describe a structure that works well across a wide range of scenarios. However, the options are virtually limitless and we expect that more effective designs will be found for more specific problems. The first layer of the neural network in our knockoff machine takes a vector of original variables $X$ and a $p$-dimensional noise vector $V \sim \mathcal{N}(0,I)$ as input. Then a collection of $h$ latent variables is produced by taking different linear combinations of the input and applying to each a nonlinear activation function. The connections in this layer are represented in the schematics of Figure~\ref{fig:network-input}, where $p=3$ and $h=5$. The same pattern of linear and nonlinear units is repeatedly applied to the hidden variables, through $K$ layers of width $h$, as shown in Figure~\ref{fig:network-hidden}. Finally, a similarly designed output layer returns a $p$-dimensional vector, as depicted in Figure~\ref{fig:network-output}. Following the approach of generative moment matching networks \cite{li2015generative}, we replaced the unbiased maximum mean discrepancy loss in \eqref{eq:J-MMD} with a slightly modified version that is always positive because it performs better in practice; see \cite[Section 4.3]{li2015generative} for technical details. In order to reduce the training time, the machines are fitted by applying stochastic gradient descent with momentum as it is customary in the field.

We have observed superior performance when a modified decorrelation penalty is adopted instead of the simpler
expression in \eqref{eq:J-power}. For this purpose, we suggest using
 \begin{align*}
   J_{ \text{decorrelation}}(\X, \tilde{\X}) 
   & = \| \text{diag}(\hat{G}_{X\tilde{X}})-1+s^*_{\text{SDP}}(\hat{G}_{XX})  \|_2^2,
 \end{align*}
where $\hat{G}$ is defined as in \eqref{eq:covariance-joint} and  $s^*_{\text{SDP}}$ is the solution to the semi-definite program
\begin{align*}
  s^*_{\text{SDP}}(\Sigma)
  = \argmin_{s \in [0,1]^p} & \sum_{j=1}^{p} \left| 1-s_j\right|, \\
  \text{s.t.} \qquad & 2 \Sigma \succeq \text{diag}(s).
\end{align*}
The above optimization problem is the same used in \cite{candes2016panning} to minimize the pairwise correlations between the knockoffs and the original variables, in order to boost the power, for the special case of $X \sim \Normal{0}{\Sigma}$. Under the Gaussian assumption, the constraint $2 \Sigma \succeq \text{diag}(s) \succeq 0$ is necessary and sufficient to ensure that the joint covariance matrix of $(X, \tilde{X})$ is positive semidefinite.


\section{Robustness and diagnostics} \label{sec:robust-diagnostics}
\subsection{Measuring goodness-of-fit} \label{sec:tuning}


For any fixed data source $P_X$, the goodness-of-fit of a conditional model producing approximate knockoff copies $\tilde{X} \mid X$ can be informally defined as the compatibility of the joint distribution of $(X,\tilde{X})$ with the exchangeability property in \eqref{eq:knock-2}. By defining and evaluating different measures of such discrepancy, the quality of our deep knockoff machines can be quantitatively compared to that of existing alternatives. This task is a special case of the two-sample problem mentioned in Section~\ref{sec:deep-models}, with the additional complication that a large number of distributions are to be simultaneously analyzed. In fact, for any fixed $P_X$ and $P_{\tilde{X} \mid X}$, one should verify whether all of the following null hypotheses are true:
\begin{align*}
  \mathcal{H}_0^{(j)} : P_{(X,\tilde{X})} = P_{(X, \tilde{X})_{\text{swap}(j)}},
  \qquad j \in \{1,\ldots,p\}.
\end{align*}
In order to reduce the number of comparisons, we will instead consider the following two hypotheses:
\begin{align} \label{eq:goodness-of-fit-hypotheses}
  & \mathcal{H}_0^{\text{full}} : P_{(X,\tilde{X})} = P_{(\tilde{X}, X)},
  & \mathcal{H}_0^{\text{partial}} : P_{(X,\tilde{X})} = P_{(X, \tilde{X})_{\text{swap}(S)}},
\end{align}
where $S$ is a random subset of $\{1,\ldots,p\}$, chosen uniformly
and independently of $X,\tilde{X}$, such that $j \in S$ with probability $1/2$.
Either hypothesis can be separately investigated by applying a variety of existing two-sample
tests, as described below. In order to study
$\mathcal{H}_0^{\text{full}}$, we define $\Z_1$ and $\Z_2$ as two
independent sets of $n$ observations, respectively drawn from the
distribution of $Z_1 = (X,\tilde{X})$ and $Z_2 = (\tilde{X}, X)$. The
analogous tests of $\mathcal{H}_0^{\text{partial}}$ can be performed
by defining $\Z_2$ as the family of samples $(X,
\tilde{X})_{\text{swap}(S)}$, and they are omitted in the interest of
space. 

\textbf{Covariance diagnostics.} It is natural to begin with a comparison of the covariance matrices of
$Z_1$ and $Z_2$, namely $G_1,G_2 \in \RR^{2p \times 2p}$. For this
purpose, we compute the following statistic meant to test the
hypothesis that $G_1=G_2$:
\begin{align} \label{eq:metric-covariance}
   \widehat{\varphi}_\text{COV}
  & = \frac{1}{n(n-1)} \sum_{i=1}^{n} \sum_{j \neq i}^n \left[ (Z_{1i}^\top Z_{1j})^2 + (Z_{2i}^\top Z_{2j})^2 \right] - 
    \frac{2}{n^2} \sum_{i=1}^{n} \sum_{j=1}^{n} (Z_{1i}^\top Z_{2j})^2.
\end{align}
This quantity is an unbiased estimate of $\|G_1-G_2\|_F^2 = \text{Tr}(G_1^{\top}G_1) + \text{Tr}(G_2^{\top}G_2) - 2\text{Tr}(G_1^{\top}G_2)$, if $Z_1$ and $Z_2$ have zero mean \cite{li2012two}. In practice, $Z_1$ and $Z_2$ will be centered if this assumption does not hold. 
The asymptotic distribution of \eqref{eq:metric-covariance} can be derived under mild conditions, thus yielding a non-parametric test of the null hypothesis that $G_1=G_2$ \cite{li2012two}. However, since our goal is to compare knockoffs generated by alternative algorithms, we will simply interpret larger values of \eqref{eq:metric-covariance} as evidence of a worse fit. 

\textbf{MMD diagnostics.} While being intuitive and easy to evaluate, the above diagnostic is
limited as it does not capture the higher-order moments of $(X,
\tilde{X})$. 
Therefore, different diagnostics should be used in order to have power
against other alternatives. 
A natural choice is to rely on the maximum mean discrepancy, on which
the construction of the deep knockoff machines in
Section~\ref{sec:machines-higher-order} is based. In particular, the
first null hypothesis in \eqref{eq:goodness-of-fit-hypotheses} can be
tested by computing
\begin{align} \label{eq:metric-MMD}
  & \widehat{\varphi}_\text{MMD} = \widehat{\mathcal{D}}_{\text{MMD}} \left( \Z_1, \Z_2 \right),
\end{align}
where the function $\widehat{\mathcal{D}}_{\text{MMD}}$ is defined as
in \eqref{eq:MMD-hat}. See  \cite{gretton2012kernel} for details. Since this is an unbiased estimate of the maximum mean discrepancy between the two distributions, large values can again be interpreted as evidence against the null. On the other hand, exact knockoffs will lead to values equal to zero on average. 

\textbf{KNN diagnostics.} The $k$-nearest neighbors test \cite{schilling1986multivariate} can also be employed to obtain a non-parametric measure of goodness-of-fit. For simplicity, we consider here the special case of $k=1$. For each sample $z_{li} \in \Z_l$, with $l\ \in \{1,2\}$, we denote the nearest neighbor in Euclidean distance of $z_{li}$ among $\Z = \Z_1 \cup \Z_2 \setminus \{z_{li}\}$ as $N\!N(z_{li})$. Then, we define $I_{l}(i)$ to be equal to one if $N\!N(z_{li}) \in \Z_l$ and zero otherwise, and compute
\begin{align}\label{eq:metric-KNN}
  \widehat{\varphi}_\text{KNN}
  & = \frac{1}{2n} \sum_{i=1}^{n} \left[ I_1(i) + I_2(i) \right].
\end{align}
This quantity is the fraction of samples whose nearest neighbor happens to originate from the same distribution. In expectation, $\widehat{\varphi}_\text{KNN}$ is equal to $1/2$ if the two distributions are identical, while larger values provide evidence against the null \cite{schilling1986multivariate}. A rigorous test can be performed to determine whether to reject any given knockoff generator, by applying the asymptotic significance threshold derived in \cite{schilling1986multivariate}. However, since exact knockoffs may be difficult to achieve in practice, we choose to use these statistics to grade the quality of different approximations. According to this criterion one should prefer knockoff constructions leading to values of this statistic that are closer to $1/2$. 

\textbf{Energy diagnostics.} Finally, the hypotheses in \eqref{eq:goodness-of-fit-hypotheses} can also be tested in terms of the energy distance \cite{szekely2013energy}, defined as
\begin{align}
\mathcal{D}_{ \text{Energy}}(P_{Z_1},P_{Z_2})
& = 2\EE_{Z_1,Z_2} \left[ \|Z_1 - Z_2\|_2\right] - \EE_{Z_1,Z_1'} \left[ \|Z_1 - Z_1' \|_2\right] - \EE_{Z_2,Z_2'} \left[\|Z_2-Z_2'\|_2\right],
\end{align}
where $Z_1, Z_1', Z_2, Z_2'$ are independent samples drawn from $P_{Z_1}$ and $P_{Z_2}$, respectively. Assuming finite second moments, one can conclude that $\mathcal{D}_{ \text{Energy}} \geq0$, with equality if and only if $Z_1$ and $Z_2$ are identically distributed \cite{szekely2013energy}. Therefore, we follow the approach of \cite{szekely2013energy} and define the empirical estimator
\begin{align*}
\widehat{\mathcal{D}}_{\text{Energy}} \left( \Z_1, \Z_2 \right)
& = \frac{2}{n^2} \sum_{i=1}^{n}\sum_{j=1}^{n} \|Z_{1i}-Z_{2j}\|_2 - \frac{1}{n^2} \sum_{i=1}^{n}\sum_{j=1}^{n} \|Z_{1i}-Z_{1j}\|_2 - \frac{1}{n^2} \sum_{i=1}^{n}\sum_{j=1}^{n} \|Z_{2i}-Z_{2j}\|_2,
\end{align*}
and the test statistic
\begin{align} \label{eq:metric-energy}
\widehat{\varphi}_\text{Energy}
& = \frac{n}{2}\widehat{\mathcal{D}}_{\text{Energy}} \left( \Z_1, \Z_2 \right).
\end{align}
The quantity in \eqref{eq:metric-energy} can be shown to be always positive, and it leads to a consistent test for the equality in distribution \cite{szekely2013energy}, under the assumption of finite second moments. More precisely, this statistic converges in probability to $\E{\|Z_1-Z_2\|_2}$ as the sample size $n$ grows, while diverging otherwise. Therefore, we can interpret larger values of $\widehat{\varphi}_\text{Energy}$ as evidence of a poorer fit. 

The diagnostics defined above provide a systematic way of comparing
different knockoff constructions. Sampling $\tilde{X} \mid X$ in
compliance with \eqref{eq:knock-2} for any fixed data distribution
$P_X$ is a difficult problem.
Even though the effort is motivated by the ultimate goal of performing controlled
variable selection, here the challenge is greater because even roughly
approximated knockoffs may sometimes happen to allow control of the
rate of false positives, while failing to pass the above tests. By
contrast, respecting \eqref{eq:knock-2} guarantees that the inference
will be valid. In the experiments of Section~\ref{sec:experiments} we
will show that deep machines can almost match the quality of
knockoffs produced by the existing specialized algorithms when prior information on $P_X$ 
is known, while greatly surpassing them in other cases. 

\subsection{False discovery rate under model misspecification} \label{sec:robustness}

The quality of knockoffs produced by our deep machines has been tested
according to stringent measures of discrepancy with the original data.
However, even when $(X,\tilde{X})$ is far from respecting the
exchangeability in \eqref{eq:knock-2}, the false discovery rate may
sometimes be controlled in practice. Since a scientist aiming to
perform inference on real problems cannot blindly trust any
statistical method, it is important to develop a richer set of
validation tools. The strategy originally proposed in
\cite{candes2016panning} consists of making controlled numerical
experiments that replicate the model misspecification present in the
data of interest. The main idea is to sample artificial response
variables $Y$ from some known conditional likelihood given the real
explanatory variables $X$. Meanwhile, approximate knockoff copies are
generated using the best available algorithm. Since the true null
hypotheses are known in this setting, the proportion of false
discoveries can be computed after applying the knockoff filter. By
repeating this experiment a sufficient number of times, it is possible
to verify whether the false discovery rate is contained below the
nominal level. These experiments help confirm whether the knockoffs
can be applied because the distribution of $(X,\tilde{X})$ is the same
as in the real data.

\section{Numerical experiments} \label{sec:experiments}
\subsection{Experimental setup} \label{sec:experiments-setup}

The deep knockoff machine presented in Section~\ref{sec:machines} has been implemented in Python using the PyTorch library, following the design outlined in Section~\ref{sec:machine-details}. The activation units in each layer of the neural network sketched in Figure~\ref{fig:network-input} are the parametric rectified linear unit functions \cite{he2015delving}. Between the latter and the linear combinations, an additional batch normalization function \cite{ioffe2015batch} is included. The width $h$ of the hidden layers should in general depend on the dimension $p$ of $X$, and the guideline $h=10p$ works well in practice. Six such layers are interposed between the input and the output of the machine, each parametrized by different weight matrices and biases. The maximum mean discrepancy loss function is evaluated using the Gaussian mixture kernel $k(X,X') = \frac{1}{8}\sum_{i=1}^8 \exp [-\|X-X'\|_2^2 / (2\xi_i^2) ]$, with $\xi = (1, 2, 4, 8, 16, 32, 64, 128)$.

The performance of the knockoff machines is analyzed in a variety of experiments for different choices of the data distribution $P_X$, the results of which are separately presented in the subsections below. In each case the machines are trained on synthetic data sets containing $n=10^4$ realizations of $X \in \RR^{p}$, with $p=100$. Stochastic gradient descent is applied with mini-batches of size $n/4$ and learning rate $\mu=0.001$, for a total number of gradients steps $T = 10^5$. A few different values of the hyperparameters in Algorithm \ref{alg:machine-training} are considered, in the proximity of $(\gamma,\lambda,\delta)=(1,1,1)$. The performance of the deep knockoff machine is typically not very sensitive to this choice, although we will discuss how different ratios work better with certain distributions.
Upon completion of training, the goodness-of-fit of the machines is
quantified in terms of the metrics defined in
Section~\ref{sec:tuning}, namely the matching of second moments
\eqref{eq:metric-covariance}, the maximum mean discrepancy score
\eqref{eq:metric-MMD}, the $k$-nearest neighbors test
\eqref{eq:metric-KNN} with $k=1$ and the energy test
\eqref{eq:metric-energy}. These measures are evaluated on knockoff
copies generated for 1000 previously unseen independent samples drawn
from the same distribution $P_X$. The diagnostics obtained with deep
knockoff machines are compared against those corresponding to other
existing algorithms. A natural benchmark in all scenarios exposed below
 is the original second-order method in \cite{candes2016panning}, 
which is applied by relying on the empirical  covariance matrix $\hat{\Sigma}$ 
computed on the same data used  to train the deep machine. 
Moreover, we also consider exact knockoff
constructions with perfect oracle knowledge of $P_X$ as ideal competitors.

Finally, numerical experiments are carried
out by performing variable selection in a controlled setting, where
the response is simulated from a known conditional likelihood. 
For each sample $i \in \{1,\ldots,m\}$, 
the response variable $Y^i \in \RR$ is sampled according to
$Y^i \sim \mathcal{N}(X^i\beta,1)$, with $\beta \in \RR^p$ containing $30$
randomly chosen non-zero elements equal to $a / \sqrt{m}$.
The experiments are repeated 1000 times, for different
values of the signal amplitude $a$ and the number of observations
$m$. The importance measures are defined by fitting the elastic-net
\cite{zou2005regularization} on the augmented data matrix
$[\X, \tilde{\X}] \in \reals^{m \times 2p}$ and $\Y \in \reals^m$. More precisely, we compute
$(\hat{\beta} , \tilde{\beta}) \in \RR^{2p}$ as
\begin{align} \label{eq:knock-stats}
  (\hat{\beta} , \tilde{\beta})
  & = \argmin_{(b,\tilde{b})} \left\{ \frac{1}{m} \|\Y - \X b -\tilde{\X} \tilde{b}\|_2^2 + (1-\alpha) \frac{\tau}{2} \left(\|b\|_2^2 + \|\tilde{b}\|_2^2 \right) + \alpha \tau \left(\|b\|_1 + \|\tilde{b}\|_1 \right) \right\},
\end{align}
with the value of $\tau$ tuned by 10-fold cross validation and some fixed $\alpha \in [0,1]$. The knockoff filter is applied on the statistics $W_j = |\hat{\beta}_j| - |\tilde{\beta}_j| $, for all $1 \leq j \leq p$, at the nominal level $q=0.1$. The power and the false discovery rate corresponding to knockoffs generated by different algorithms can be evaluated and contrasted, as a consequence of the exact knowledge of the ground truth. It is important to stress that these experiments and all the diagnostics described above only rely on new observations from $P_X$, generated independently of those on which the machine is trained. 

\subsection{Multivariate Gaussian} \label{sec:experiment-gaussian}

\begin{figure}[!htb]
  \centering
    \includegraphics[width=0.6\textwidth]{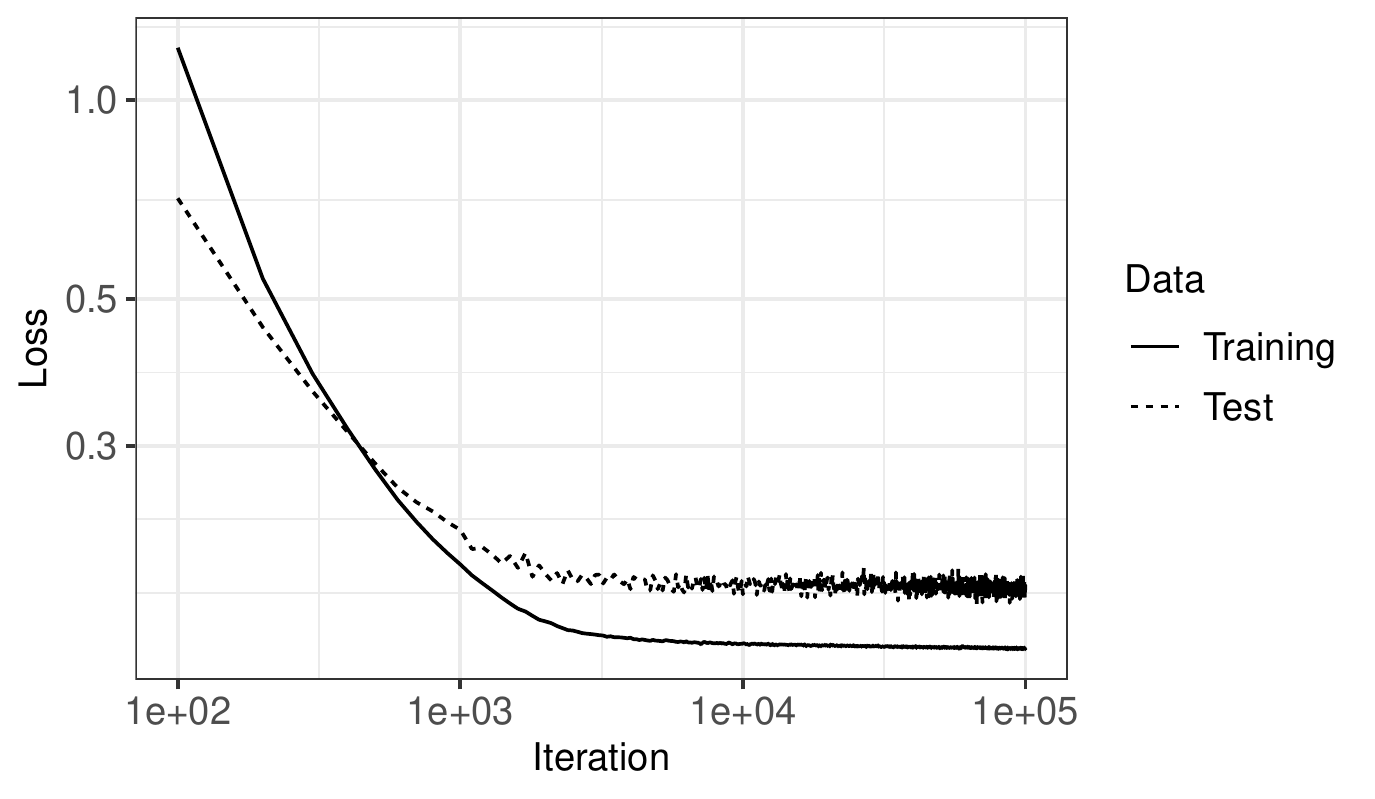}
  \caption{Evolution of the objective function for a deep machine while learning to generate knockoffs for multivariate Gaussian variables. The continuous line shows the loss \eqref{eq:J-loss} on the training set, while the dashed line indicates the loss evaluated on an independent test set.} \label{fig:sim-gaussian-progress}
\end{figure}

The first example that we present concerns the multivariate Gaussian
distribution, for which the exact construction of knockoffs in
\cite{candes2016panning} provides the ideal benchmark. For simplicity
we consider $P_X$ to be an autoregressive process of order one, with
correlation parameter $\rho=0.5$, such that
$X \sim \mathcal{N}(0,\Sigma)$ and $\Sigma_{i,j} = \rho^{|i-j|}$. A
deep knockoff machine is trained with hyperparameters
$(\gamma,\lambda,\delta)=(1,1,1)$ and the value of its loss
\eqref{eq:J-loss} is plotted in Figure \ref{fig:sim-gaussian-progress} as a function of
the training time. 

\begin{figure}[!htb]
    \centering
    \begin{subfigure}[t]{0.49\textwidth}
        \centering
        \includegraphics[width=0.8\textwidth]{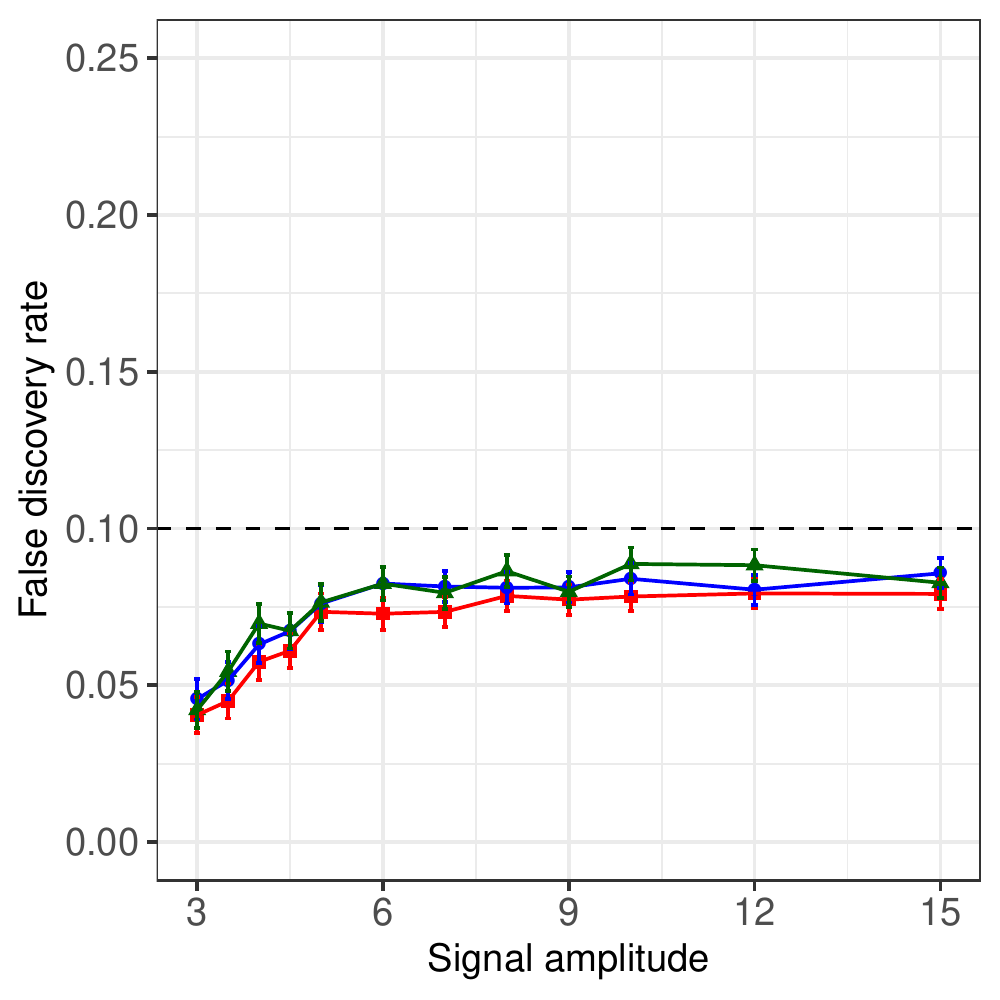}
        \caption{}  \label{fig:sim-gaussian-test-fdr}
    \end{subfigure}
    ~
    \begin{subfigure}[t]{0.49\textwidth}
        \centering
        \includegraphics[width=0.8\textwidth]{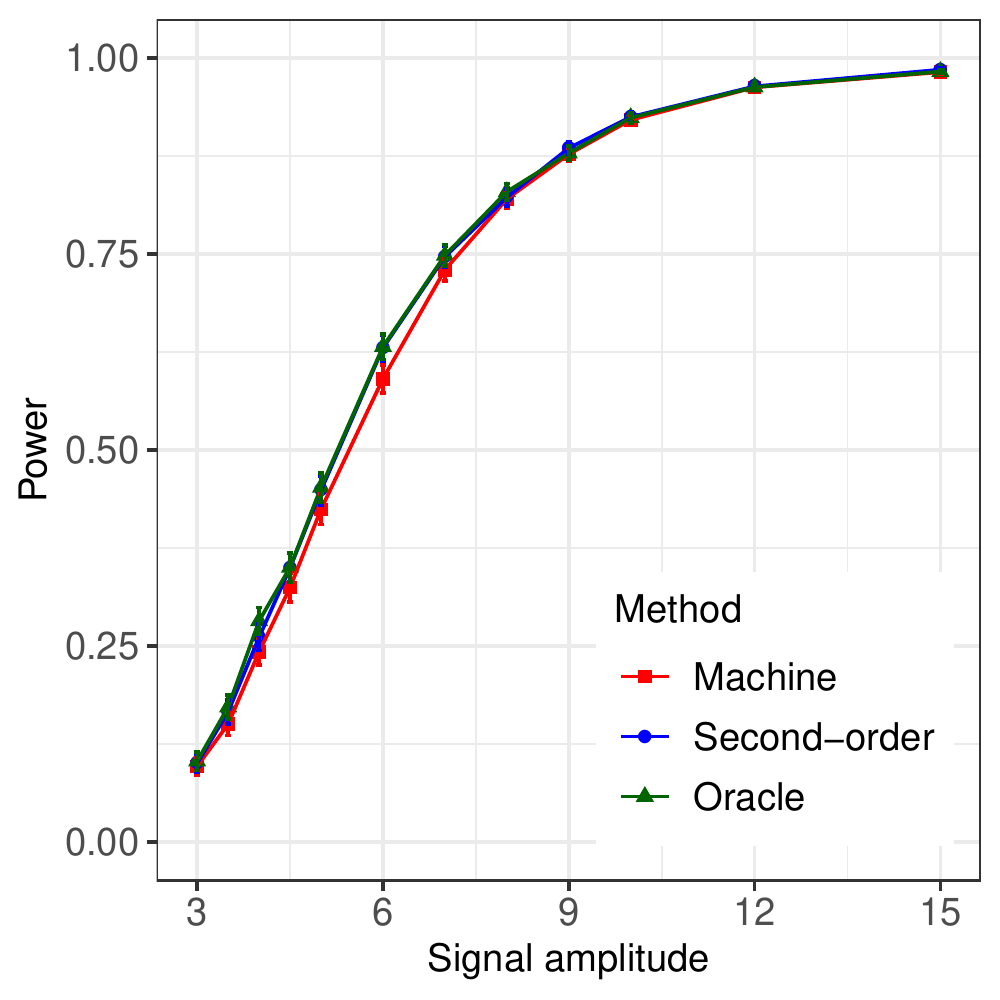}
        \caption{}  \label{fig:sim-gaussian-test-pow}
    \end{subfigure}
  \caption{Numerical experiments with multivariate Gaussian variables and simulated response. The performance of the machine (red) is compared to that of second-order (blue) and oracle (green) knockoffs. The false discovery rate (a) and the power (b) are computed by averaging over 1000 independent experiments. The three curves in (b) are almost indistinguishably overlapping. } \label{fig:sim-gaussian-test}
\end{figure}

\begin{figure}[!htb]
  \centering
    \includegraphics[width=\textwidth]{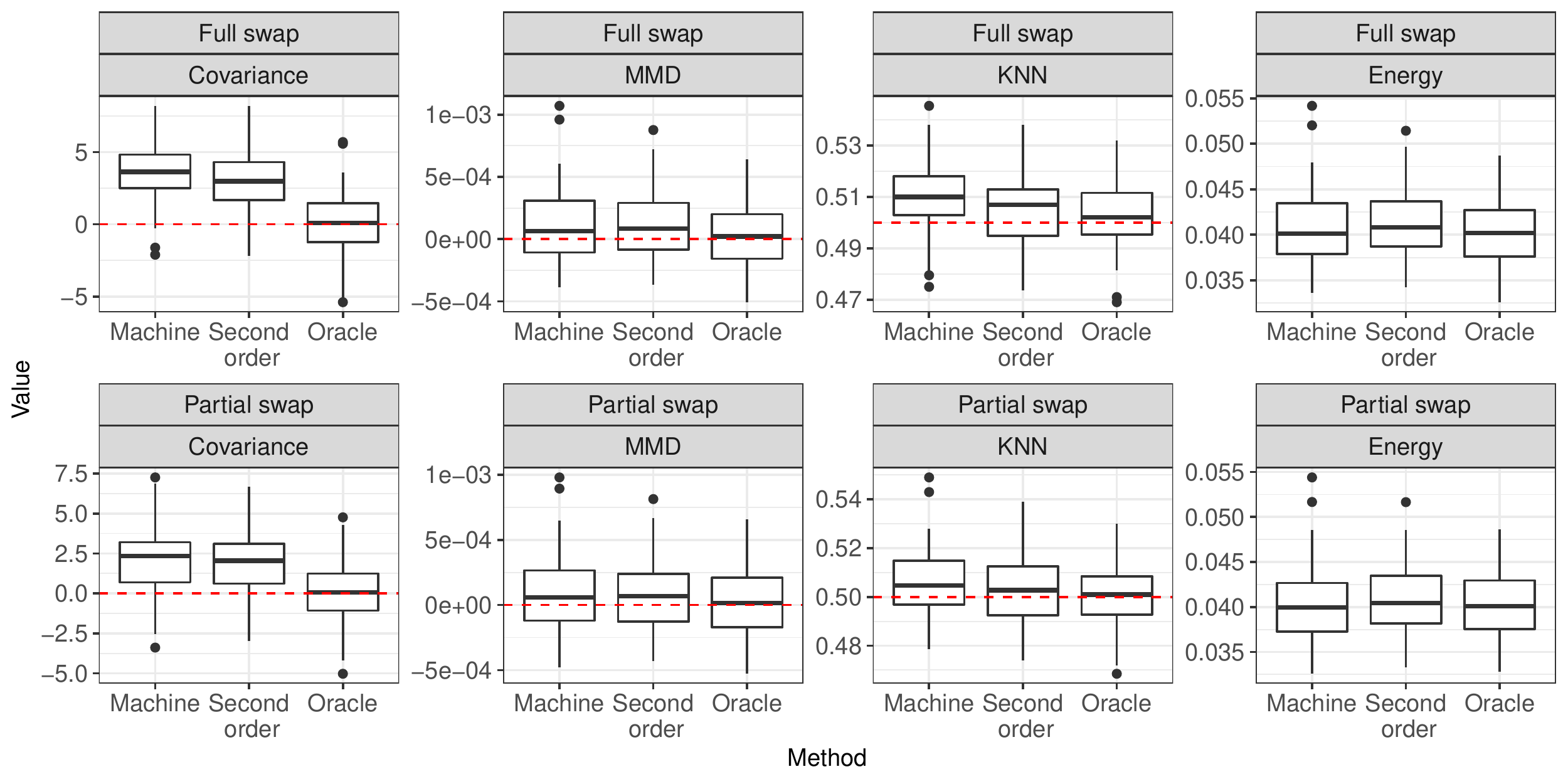}
  \caption{Boxplot comparing different knockoff goodness-of-fit diagnostics for multivariate Gaussian variables, obtained with the deep machine, the second-order method and the oracle construction, on 100 previously unseen independent datasets of size $n=1000$.} \label{fig:sim-gaussian-diagnostics}
\end{figure}

\begin{figure}[!htb]
  \centering
    \includegraphics[width=0.3\textwidth]{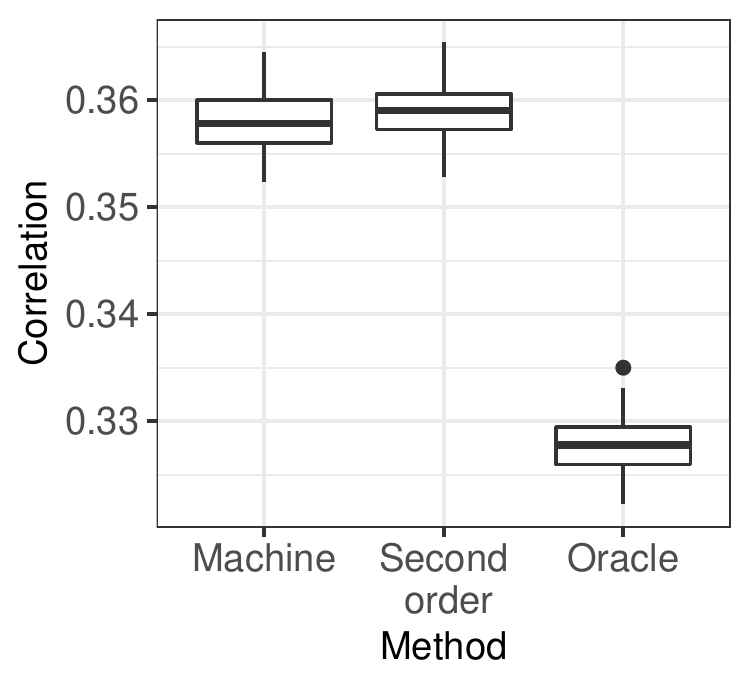}
  \caption{Boxplot comparing the average absolute pairwise correlation
    between variables and knockoffs for a multivariate Gaussian
    distribution, as in
    Figure~\ref{fig:sim-gaussian-diagnostics}. Lower values tend to
    indicate more powerful knockoffs. The numerical values on the vertical axis
    show that the differences between the three methods are not very large.} \label{fig:sim-gaussian-diagnostics-self}
\end{figure}

The controlled numerical experiments are carried out on synthetic datasets containing $m=150$ samples, and setting $\alpha=0.1$ in \eqref{eq:knock-stats}. The results corresponding to the deep machine are shown in Figure~\ref{fig:sim-gaussian-test} as a function of the signal amplitude. The performance is compared to that of the second-order method in \cite{candes2016panning} and an oracle that constructs exact knockoffs by applying the formula in \eqref{eq:machine-gaussian} with the true covariance matrix $\Sigma$.
The value of $s$ in \eqref{eq:machine-gaussian} is determined by solving the semi-definite program \cite{candes2016panning} from Section~\ref{sec:machine-details}. The goodness-of-fit of these three alternative knockoff constructions is further investigated in terms of the diagnostics defined earlier, as shown in Figure~\ref{fig:sim-gaussian-diagnostics}. Unsurprisingly, the knockoffs generated by the oracle are perfectly exchangeable, while the deep machine and the second-order knockoffs are almost equivalent. Finally, Figure~\ref{fig:sim-gaussian-diagnostics-self} suggests that the oracle has the potential to be slightly more powerful, as it can generate knockoffs with smaller pairwise correlations with their original counterparts.

\begin{figure}[!htb]
  \centering
    \includegraphics[width=0.75\textwidth]{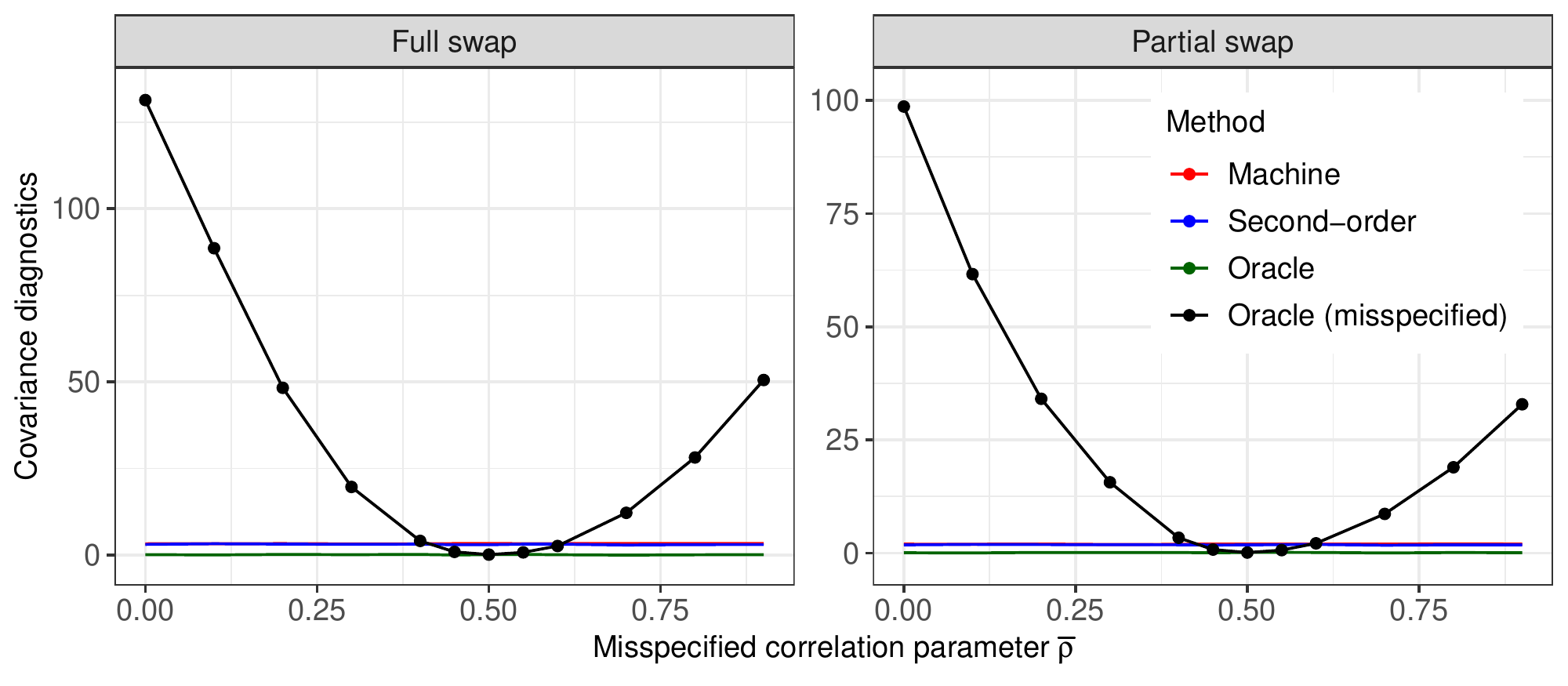}
  \caption{Covariance goodness-of-fit diagnostics for a misspecified Gaussian autoregressive knockoff oracle (black) as a function of its correlation parameter $\bar{\rho}$. These measures are compared to those of the other methods, also shown in Figure~\ref{fig:sim-gaussian-diagnostics}. The four curves indicate the expected value of the diagnostics, computed empirically on $10^6$ samples. Lower values indicate a better fit and $\bar{\rho}=0.5$ corresponds to the correct model. The lines corresponding to the deep machine and the second-order method are overlapping.} \label{fig:sim-gaussian-diagnostics-ar}
\end{figure}

The goodness-of-fit of the knockoff machine can also be measured
against that of a misguided oracle that believes $P_X$ to be an
autoregressive process of order one with correlation parameter equal
to $\bar{\rho}$. The $\tilde{X}$ thus generated are clearly not valid
knockoffs unless $\bar{\rho} = \rho$. This comparison may be helpful
because the limitations of the imperfect oracle are simpler to
understand. For example, as $\bar{\rho}$ approaches zero, $\tilde{X}$
becomes independent of $X$ and the violation of \eqref{eq:knock-2}
should be easily detectable by our tests. In the interest of space, we
only compute the second-order diagnostics in
\eqref{eq:metric-covariance} as a function of $\bar{\rho}$, and
compare them to those in
Figure~\ref{fig:sim-gaussian-diagnostics}. The results are shown in
Figure~\ref{fig:sim-gaussian-diagnostics-ar}. The misspecified oracle
leads to a significantly poorer fit than the alternative methods,
unless $\bar{\rho}$ is very close to $\rho$. This indicates that the
second-order approximation and the deep machine are capturing the true
$P_X$ rather accurately, despite having very little prior
information. Moreover, the experiment confirms that our diagnostics
are effective at detecting invalid knockoffs.

\subsection{Hidden Markov model} \label{sec:experiment-hmm}

We now consider discrete random variables $X_j \in \{0,1,2\}$, for $j \in \{1,\ldots,p\}$, distributed according to the same hidden Markov model used in \cite{sesia2017gene} to describe genotypes. In order to make the experiment more realistic, the parameters of this model are estimated from real data, by applying the imputation software fastPHASE \cite{scheet2006fast} on a reference panel of 1011 individuals from the third phase of the International HapMap project, which is freely available from \url{https://mathgen.stats.ox.ac.uk/impute/data_download_hapmap3_r2.html}. For simplicity, we restrict our attention to $p=100$ features, corresponding to variants on chromosome one whose physical positions range between 0.758 Mb and 2.456 Mb, and whose minor allele frequency is larger than 0.1. The expectation-maximization algorithm of fastPHASE is run for 35 iterations, with the number of hidden states chosen equal to $20$, and the rest of the configuration set to the default values. New observations are then sampled from the estimated $P_X$, so that the exact knockoff construction for hidden Markov models can be used as the oracle benchmark. 

\begin{figure}[!htb]
    \centering
    \begin{subfigure}[t]{0.49\textwidth}
        \centering
        \includegraphics[width=0.8\textwidth]{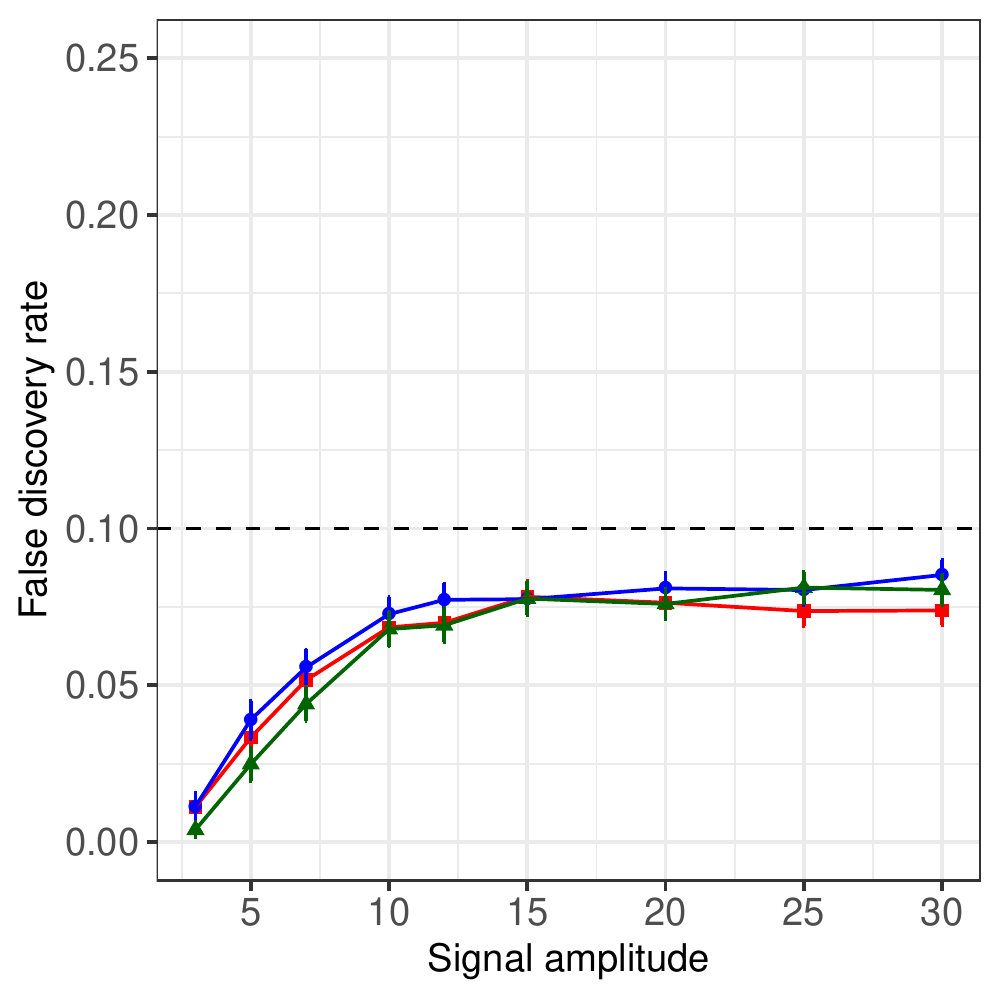}
        \caption{}  \label{fig:sim-hmm-test-fdr}
    \end{subfigure}
    ~
    \begin{subfigure}[t]{0.49\textwidth}
        \centering
        \includegraphics[width=0.8\textwidth]{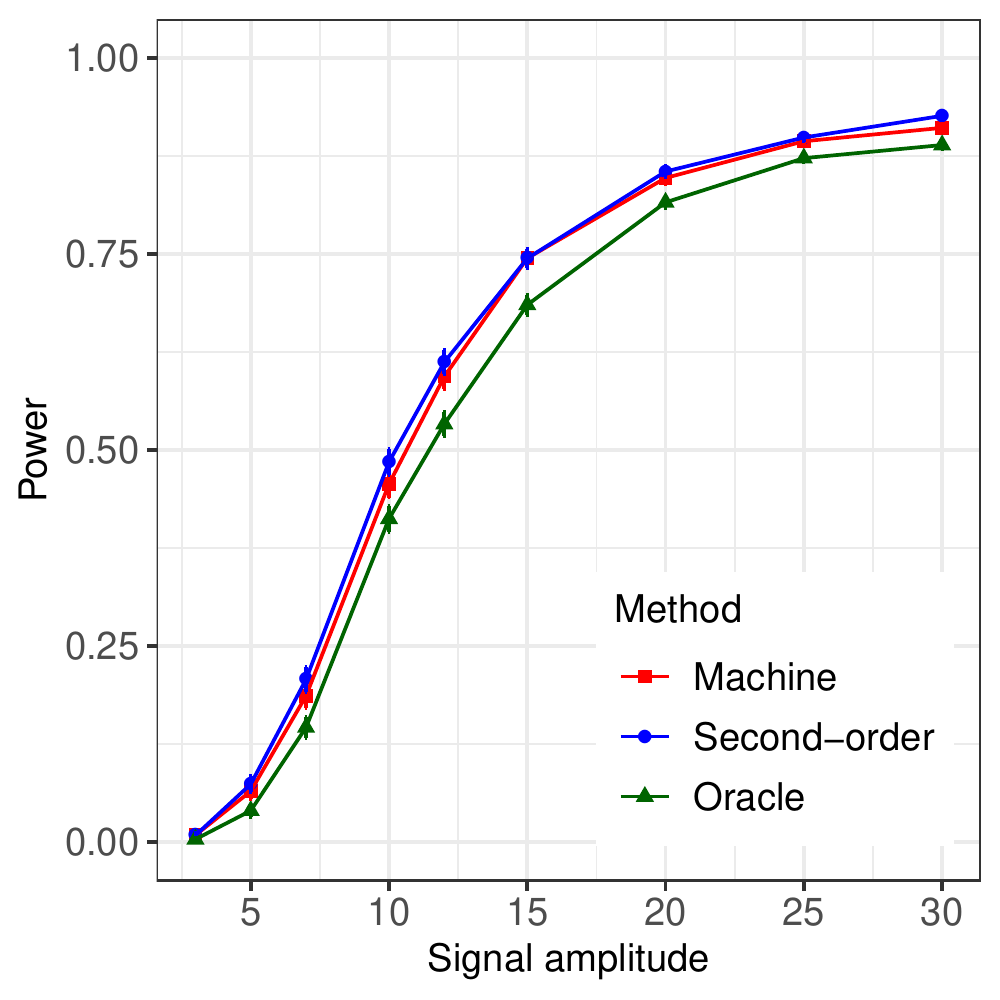}
        \caption{}  \label{fig:sim-hmm-test-pow}
    \end{subfigure}
  \caption{Numerical experiments with variables from a hidden Markov model. The other details are as in Figure~\ref{fig:sim-gaussian-test}.} \label{fig:sim-hmm-test}
\end{figure}

\begin{figure}[!htb]
  \centering
    \includegraphics[width=\textwidth]{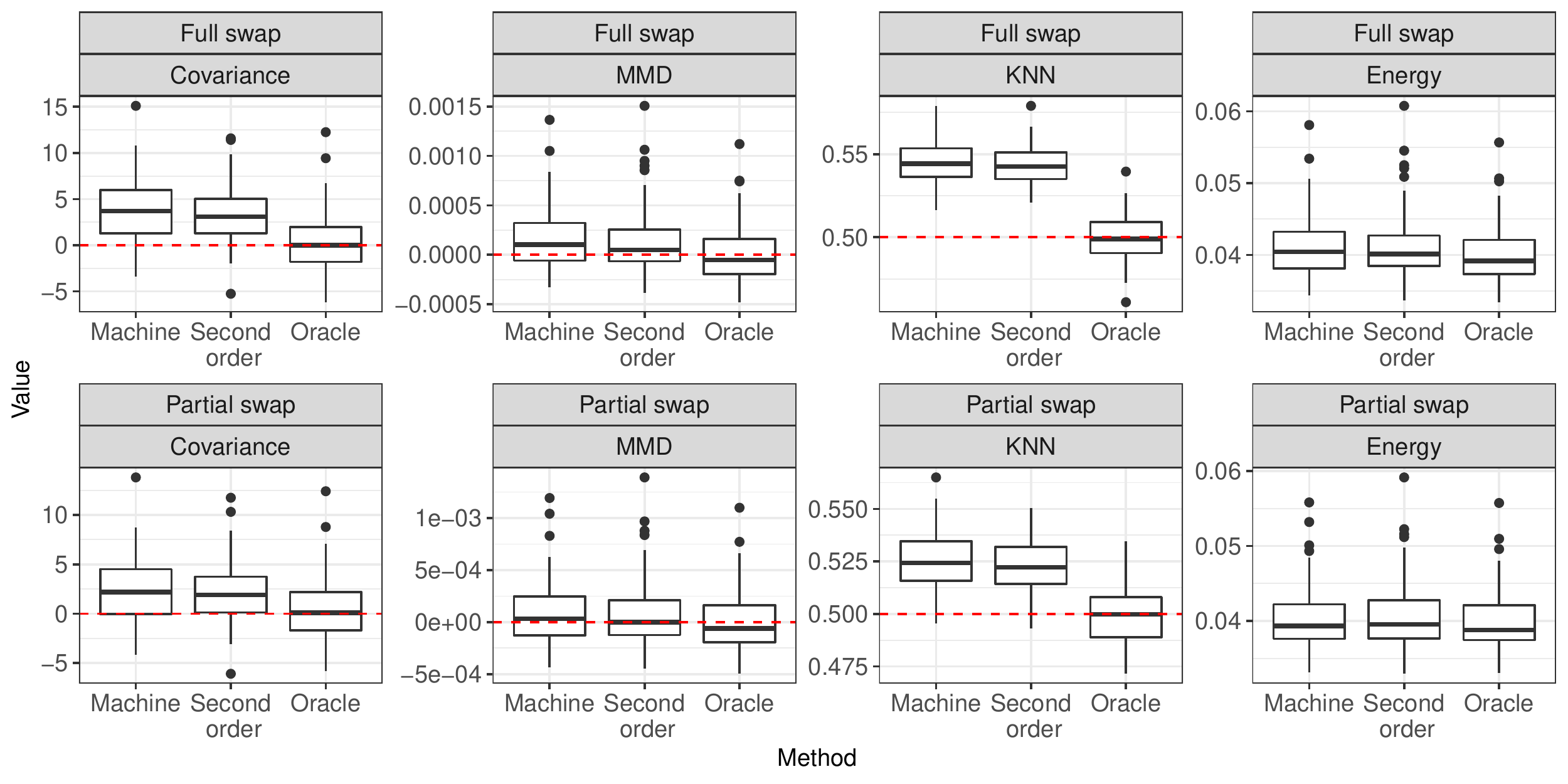}
  \caption{Boxplot comparing different knockoff diagnostics for variables sampled from a hidden Markov model. The other details are as in Figure~\ref{fig:sim-gaussian-diagnostics}.} \label{fig:sim-hmm-diagnostics}
\end{figure}

\begin{figure}[!htb]
  \centering
    \includegraphics[width=0.3\textwidth]{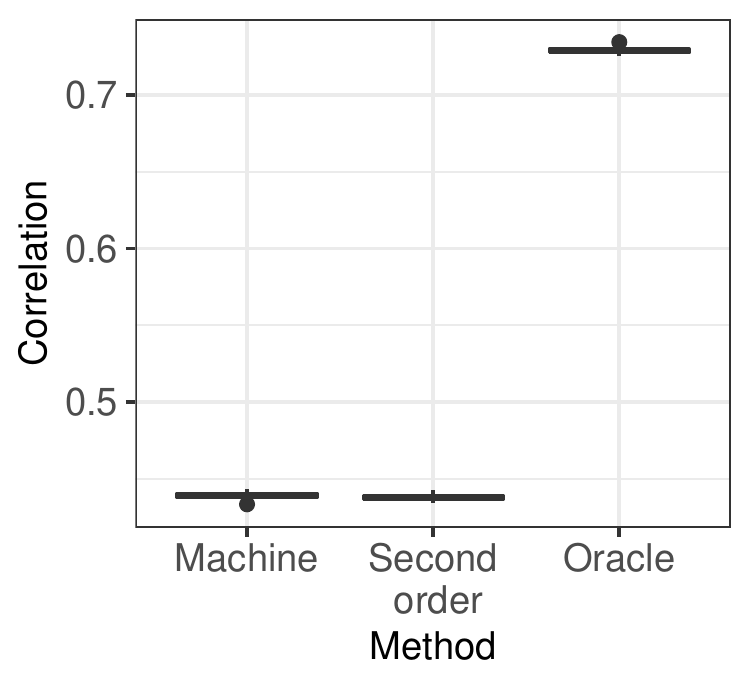}
  \caption{Boxplot comparing the average absolute pairwise correlation between variables and knockoffs for a hidden Markov model. The other details are as in Figure~\ref{fig:sim-gaussian-diagnostics-self}.} \label{fig:sim-hmm-diagnostics-self}
\end{figure}

The deep knockoff machine is trained with the hyperparameters equal to $(\gamma,\lambda,\delta)=(1,1,1)$.  The numerical experiments follow the approach outlined in Section~\ref{sec:experiments-setup}, using $m=150$ samples in each instance and setting $\alpha = 0.1$ in \eqref{eq:knock-stats}. The power and the false discovery rate are reported in Figure~\ref{fig:sim-hmm-test}, and are very similar across the three methods. However, the oracle is slightly more conservative. The goodness-of-fit diagnostics in Figures \ref{fig:sim-hmm-diagnostics} and \ref{fig:sim-hmm-diagnostics-self} indicate that the machine is almost equivalent to the second-order approximation. It may be possible to improve the performance of this deep knockoff machine by changing its architecture to account for the discrete values of this data or by tuning it more carefully.

\subsection{Gaussian mixture model} \label{sec:experiment-GMM}

\begin{figure}[!htb]
    \centering
    \begin{subfigure}[t]{0.49\textwidth}
        \centering
        \includegraphics[width=0.8\textwidth]{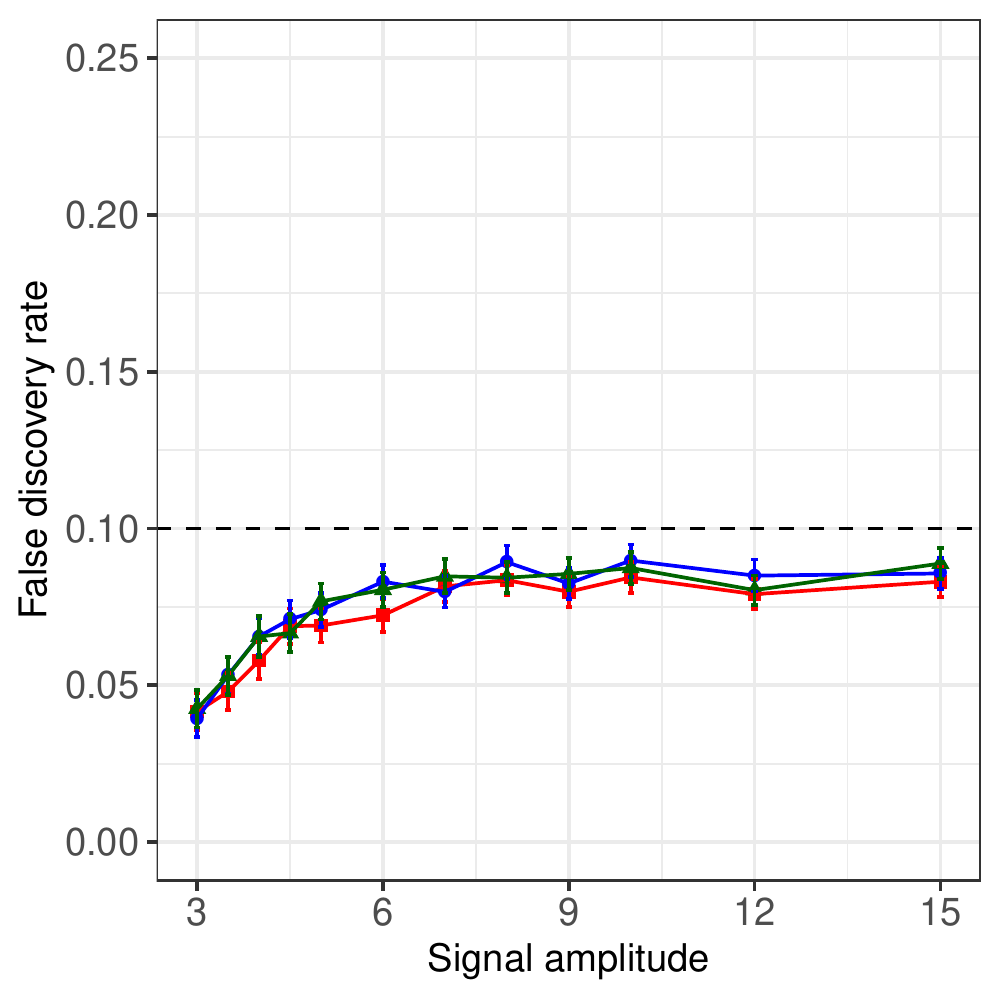}
        \caption{}  \label{fig:sim-gmm-test-fdr}
    \end{subfigure}
    ~
    \begin{subfigure}[t]{0.49\textwidth}
        \centering
        \includegraphics[width=0.8\textwidth]{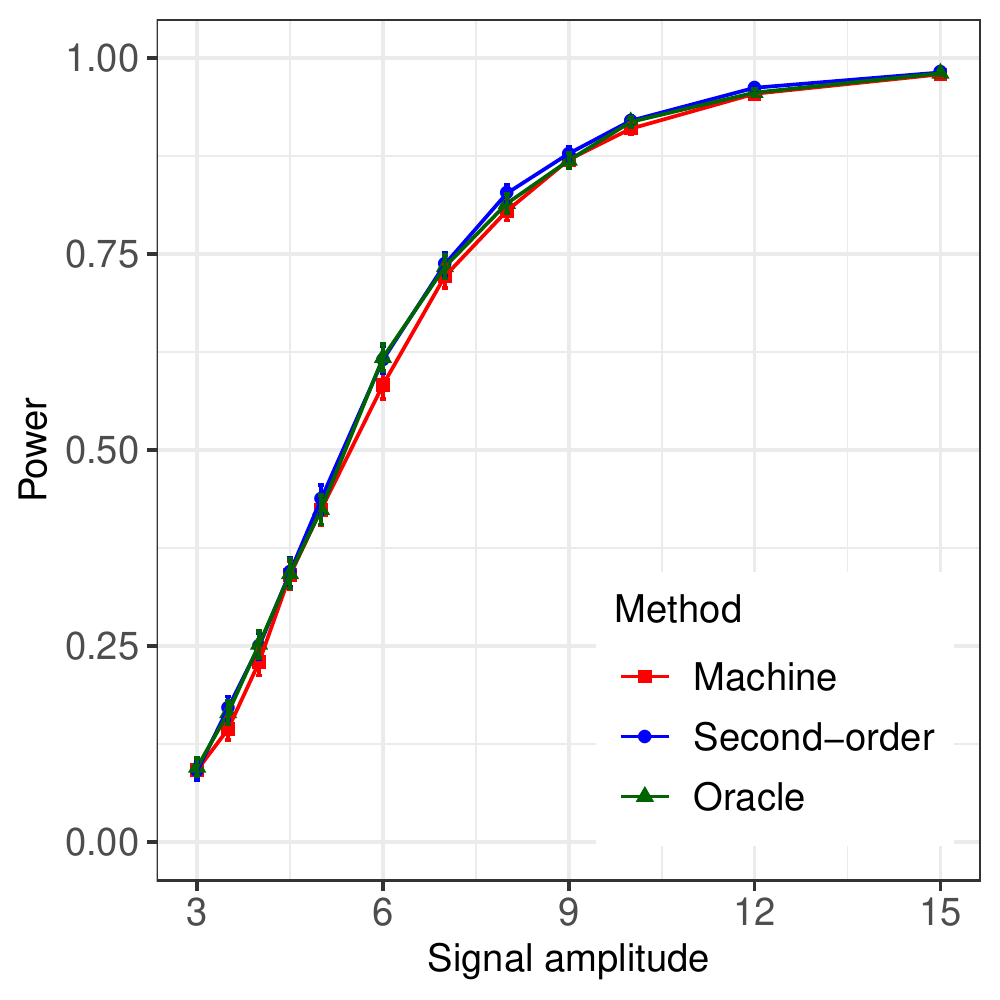}
        \caption{}  \label{fig:sim-gmm-test-pow}
    \end{subfigure}
  \caption{Numerical experiments with variables from a multivariate Gaussian mixture. The other details are as in Figure~\ref{fig:sim-gaussian-test}. The three curves in (b) are almost indistinguishably overlapping.} \label{fig:sim-gmm-test}
\end{figure}

\begin{figure}[!htb]
  \centering
    \includegraphics[width=\textwidth]{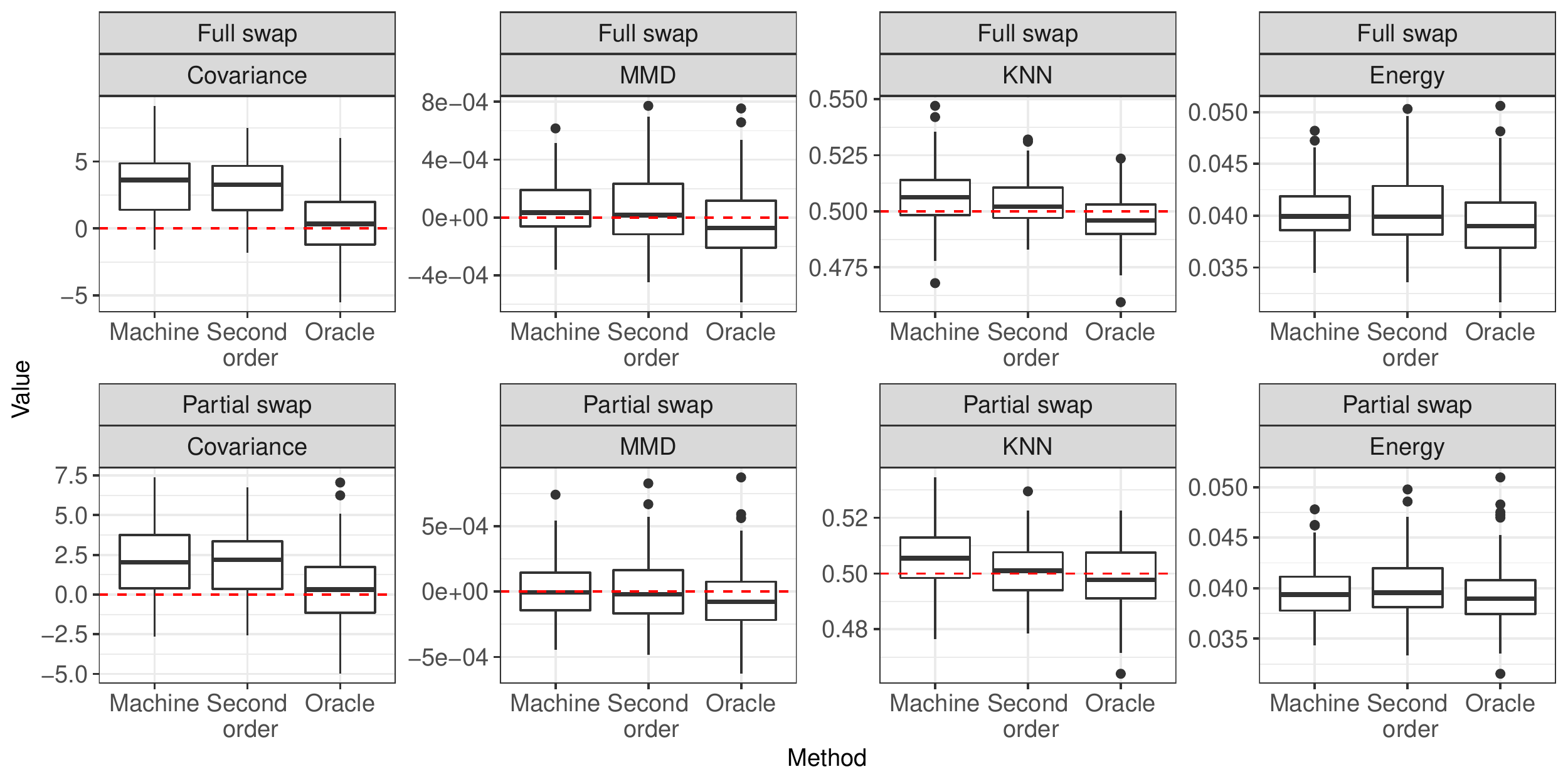}
  \caption{Boxplot comparing different knockoff diagnostics for variables sampled from a multivariate Gaussian mixture. The other details are as in Figure~\ref{fig:sim-gaussian-diagnostics}.} \label{fig:sim-gmm-diagnostics}
\end{figure}

\begin{figure}[!htb]
  \centering
    \includegraphics[width=0.3\textwidth]{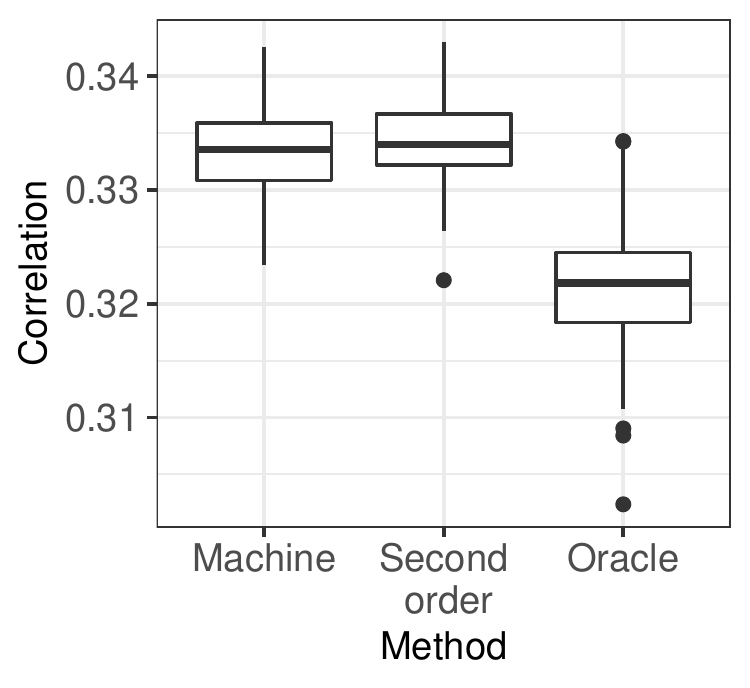}
  \caption{Boxplot comparing the average absolute pairwise correlation between variables and knockoffs for a multivariate Gaussian mixture model. The other details are as in Figure~\ref{fig:sim-gaussian-diagnostics-self}.} \label{fig:sim-gmm-diagnostics-self}
\end{figure}

The next example considers a multivariate Gaussian mixture model with equal proportions. In particular, we assume that each $X \in \RR^p$ is independently sampled from
\begin{align*}
  X \sim
  \begin{cases}
  \mathcal{N}\left(0,\Sigma_1\right), & \text{ with probability } \frac{1}{3}, \\
  \mathcal{N}\left(0,\Sigma_2\right), & \text{ with probability } \frac{1}{3}, \\ 
  \mathcal{N}\left(0,\Sigma_3\right) , & \text{ with probability } \frac{1}{3},
  \end{cases}
\end{align*}
where the covariance matrices $\Sigma_1$, $\Sigma_2$ and $\Sigma_3$
have the same autoregressive structure as in
Section~\ref{sec:experiment-gaussian}, with $\rho_1 = 0.3$,
$\rho_2 = 0.5$ and $\rho_3 = 0.7$, respectively. Exact knockoffs can be 
constructed by applying the procedure described in \cite{gimenez2018knockoffs}
to the true model parameters.
This oracle performs two simple steps. 
First, a latent mixture allocation $Z \in \{1,2,3\}$ is sampled from its posterior distribution 
given the observed $X$. Second, an exact multivariate Gaussian knockoff copy $\tilde{X}$ is 
produced conditional on $X$ and $Z$.
We then proceed with the experiments defined in
Section~\ref{sec:experiments-setup}, on $m=150$ samples and setting
$\alpha = 0.1$ in \eqref{eq:knock-stats}. The deep machine is trained
with hyperparameters equal to $(\gamma,\lambda,\delta)=(1,1,1)$. The
results of the numerical simulations, presented in
Figure~\ref{fig:sim-gmm-test}, show that the machine and the
second-order knockoffs behave as well as the oracle. 
The
goodness-of-fit diagnostics are reported in Figures
\ref{fig:sim-gmm-diagnostics} and
\ref{fig:sim-gmm-diagnostics-self}. These measures indicate that the
second-order method and the deep machine are essentially equivalent,
while nearly as accurate as the oracle.

\subsection{Multivariate Student's $t$-distribution}

In the previous experiments, deep knockoff machines matched the performance of its best competitors. At the same time, the second-order method never failed to control the false discovery rate. In contrast, the next two examples show that second-order knockoffs can indeed fail, and quite spectacularly, in the presence of different data distributions $P_X$. In particular, we now consider a multivariate Student's $t$-distribution with zero mean and $\nu = 3$ degrees of freedom, defined such that 
\begin{align*}
  X = \sqrt{\frac{\nu-2}{\nu}} \frac{Z}{\sqrt{\Gamma}},
\end{align*}
where $Z \sim \Normal{0}{\Sigma}$ and $\Gamma$ is independently drawn
from a gamma distribution with shape and rate parameters both equal to
$\nu / 2$. The covariance matrix $\Sigma$ is that of an autoregressive
process of order one with $\rho = 0.5$, as defined in
Section~\ref{sec:experiments-setup}. The scaling factor ensures that
each variable has unit variance, while their tails remain heavy. 
In fact, moments of order $\nu$ or higher are not finite.

\begin{figure}[!htb]
    \centering
    \begin{subfigure}[t]{0.49\textwidth}
        \centering
        \includegraphics[width=0.8\textwidth]{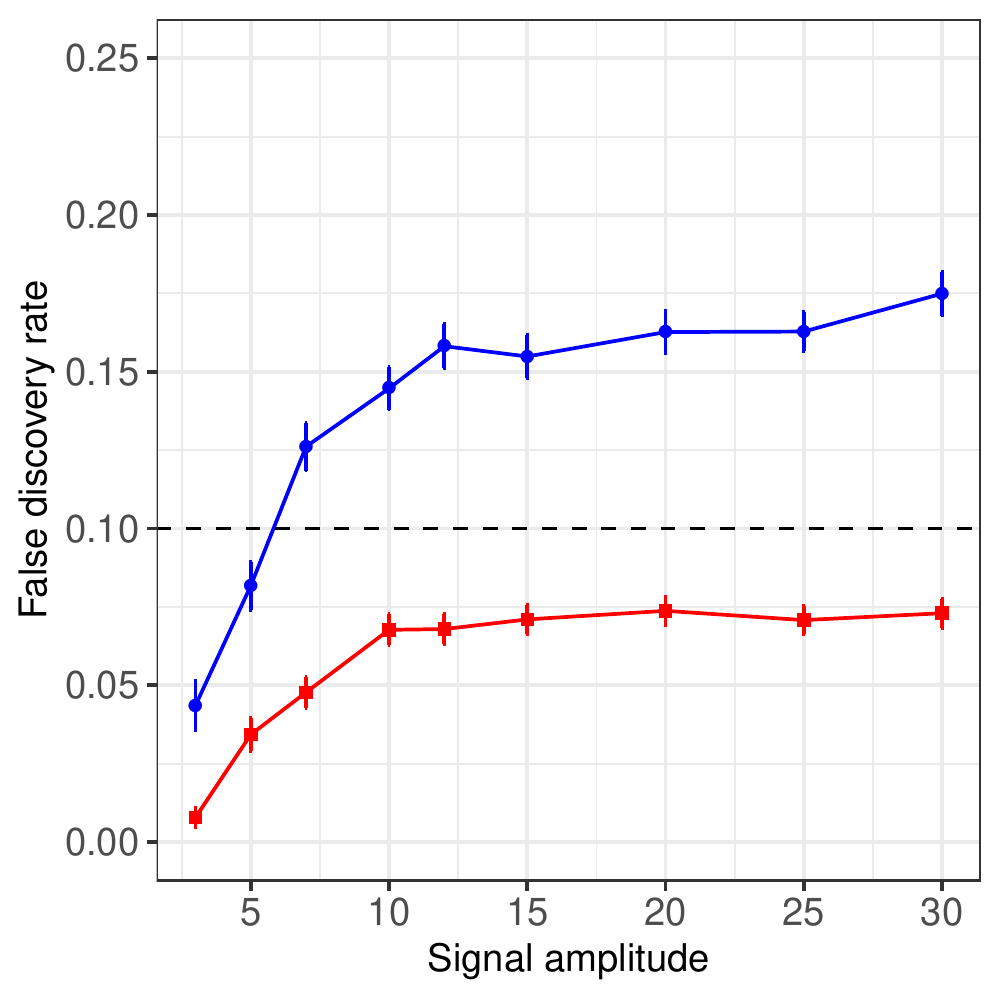}
        \caption{}  \label{fig:sim-mstudent-test-fdr}
    \end{subfigure}
    ~
    \begin{subfigure}[t]{0.49\textwidth}
        \centering
        \includegraphics[width=0.8\textwidth]{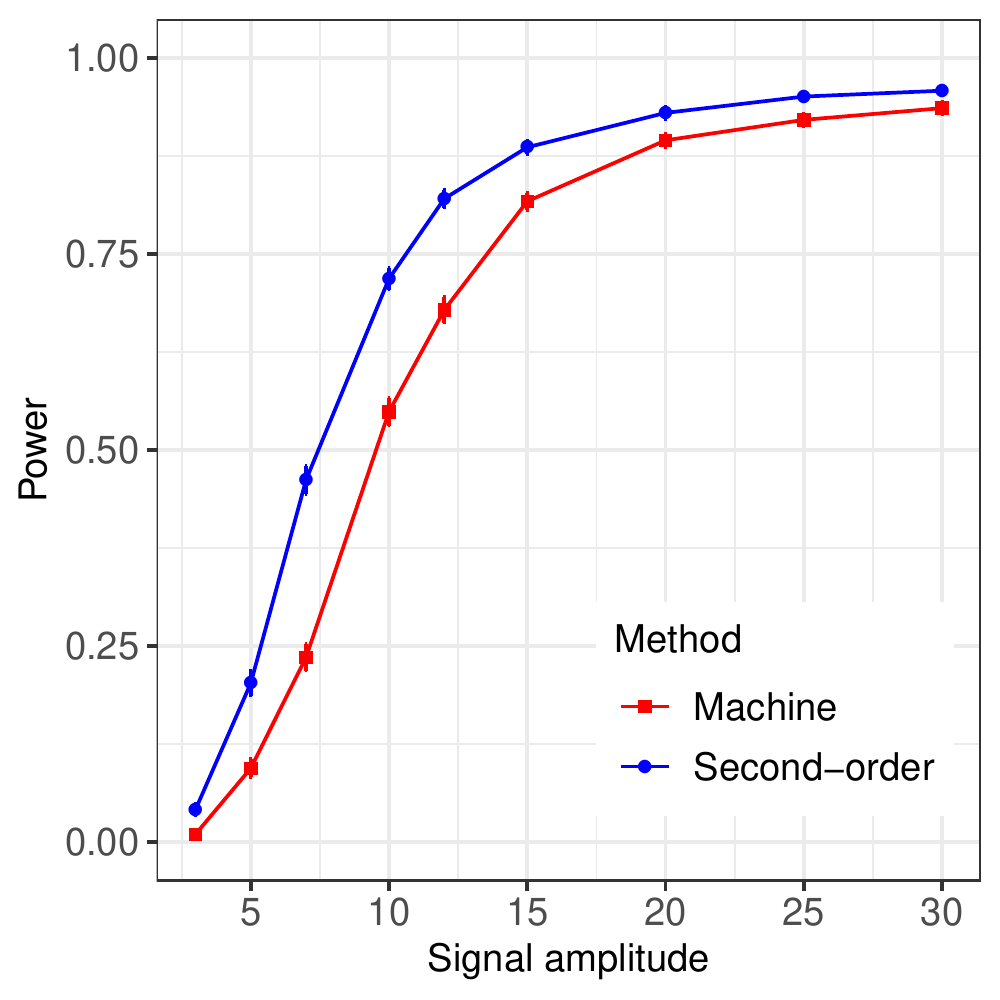}
        \caption{}  \label{fig:sim-mstudent-test-pow}
    \end{subfigure}
  \caption{Numerical experiments with a multivariate Student's $t$-distribution. The other details are as in Figure~\ref{fig:sim-gaussian-test}.} \label{fig:sim-mstudent-test}
\end{figure}

\begin{figure}[!htb]
  \centering
    \includegraphics[width=\textwidth]{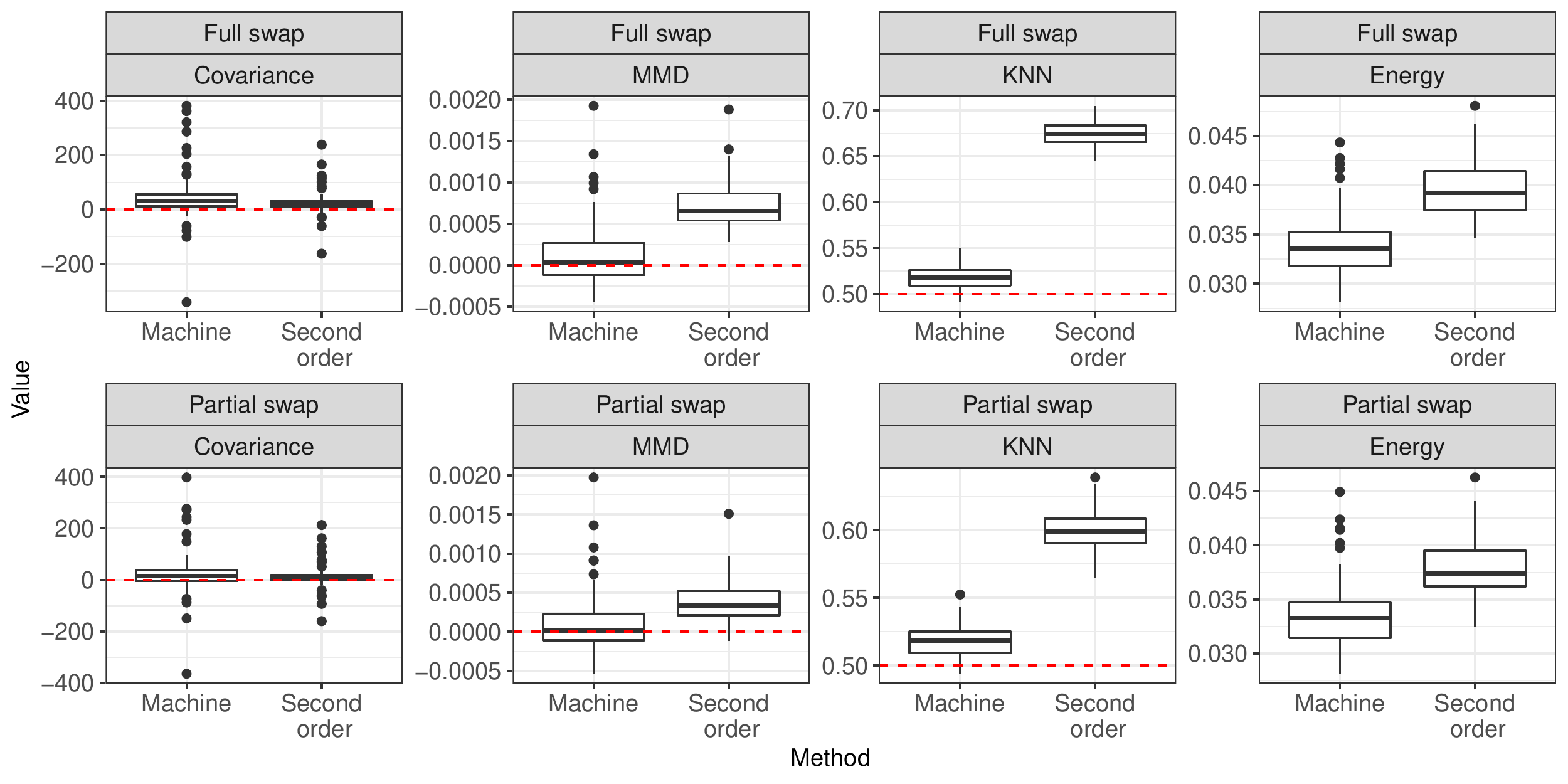}
  \caption{Boxplot comparing different knockoff diagnostics for variables sampled from a multivariate Student's $t$-distribution. The other details are as in Figure~\ref{fig:sim-gaussian-diagnostics}.} \label{fig:sim-mstudent-diagnostics}
\end{figure}

\begin{figure}[!htb]
  \centering
    \includegraphics[width=0.3\textwidth]{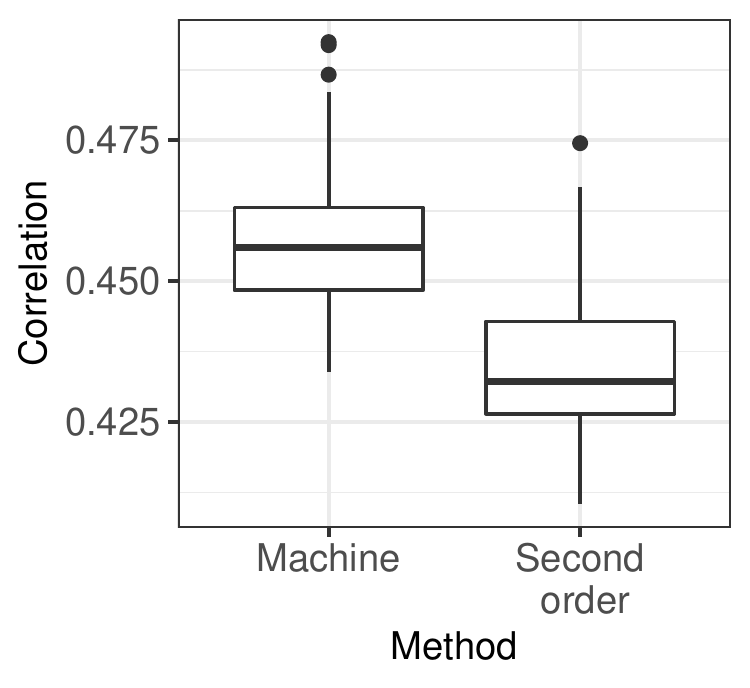}
  \caption{Boxplot comparing the average absolute pairwise correlation between variables and knockoffs for a multivariate Student's $t$-distribution. The other details are as in Figure~\ref{fig:sim-gaussian-diagnostics-self}.} \label{fig:sim-mstudent-diagnostics-self}
\end{figure}

The numerical experiments of Section~\ref{sec:experiment-gaussian} are
carried out using $m=200$ samples and setting $\alpha=0$ in
\eqref{eq:knock-stats}. The performance of a deep machine
is only compared to that of the second-order method. 
An oracle for this $P_X$ is not considered here because it is not well known, although it can be derived.
The deep machine is trained with the hyperparameters $(\gamma,\lambda,\delta)=(1,0.01,0.01)$ 
because we expect that less weight should be given to the empirical covariance matrix, which is less reliable than those in the previous experiments.
The results shown in Figure~\ref{fig:sim-mstudent-test} indicate that the deep knockoffs control the false discovery rate while second-order knockoffs fail. The goodness-of-fit diagnostics are reported in Figures \ref{fig:sim-mstudent-diagnostics} and \ref{fig:sim-mstudent-diagnostics-self}, illustrating that the deep machine significantly outperforms the second-order knockoffs.

\subsection{Sparse Gaussian variables}

\begin{figure}[!htb]
    \centering
    \begin{subfigure}[t]{0.49\textwidth}
        \centering
        \includegraphics[width=0.8\textwidth]{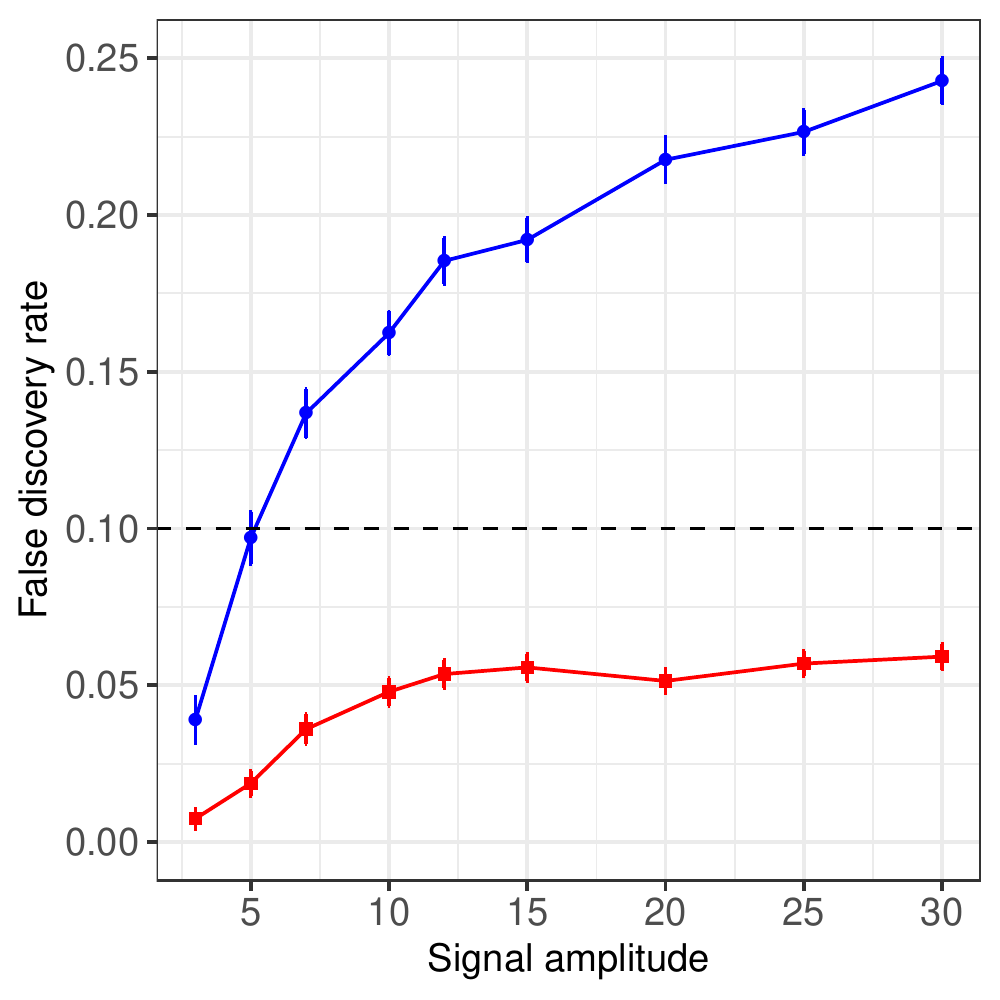}
        \caption{}  \label{fig:sim-sparse-test-fdr}
    \end{subfigure}
    ~
    \begin{subfigure}[t]{0.49\textwidth}
        \centering
        \includegraphics[width=0.8\textwidth]{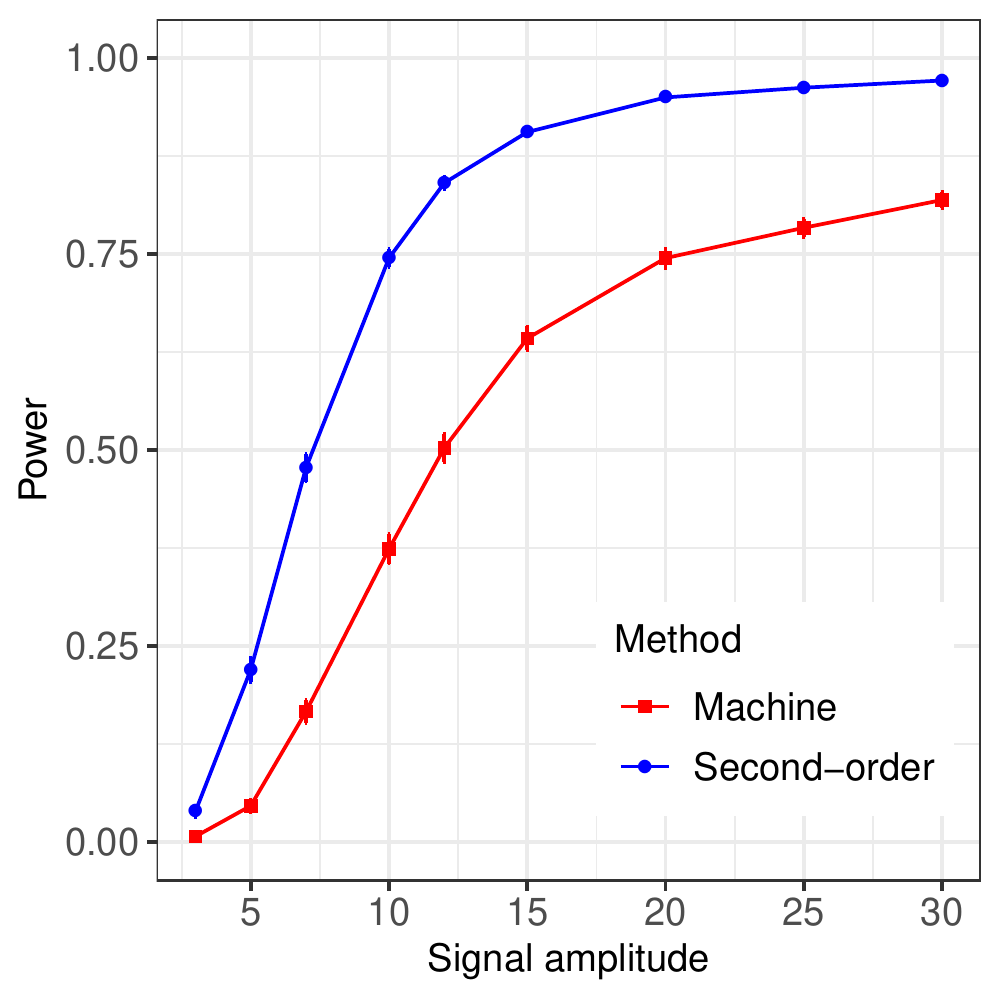}
        \caption{}  \label{fig:sim-sparse-test-pow}
    \end{subfigure}
  \caption{Numerical experiments with a sparse multivariate Gaussian distribution. The other details are as in Figure~\ref{fig:sim-gaussian-test}.} \label{fig:sim-sparse-test}
\end{figure}

\begin{figure}[!htb]
  \centering
    \includegraphics[width=\textwidth]{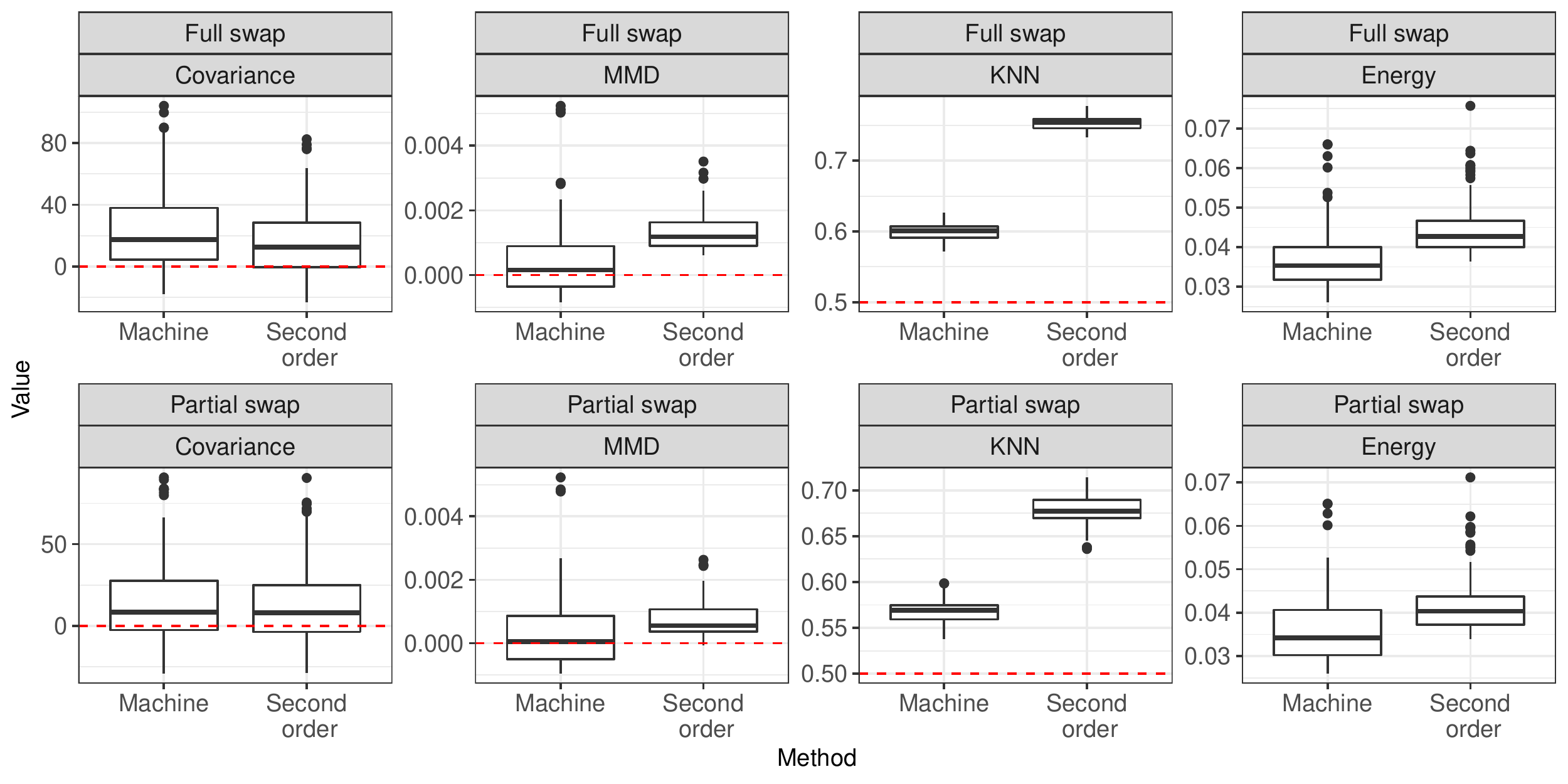}
  \caption{Boxplot comparing different knockoff diagnostics for variables sampled from a sparse multivariate Gaussian distribution. The other details are as in Figure~\ref{fig:sim-gaussian-diagnostics}.} \label{fig:sim-sparse-diagnostics}
\end{figure}

\begin{figure}[!htb]
  \centering
    \includegraphics[width=0.3\textwidth]{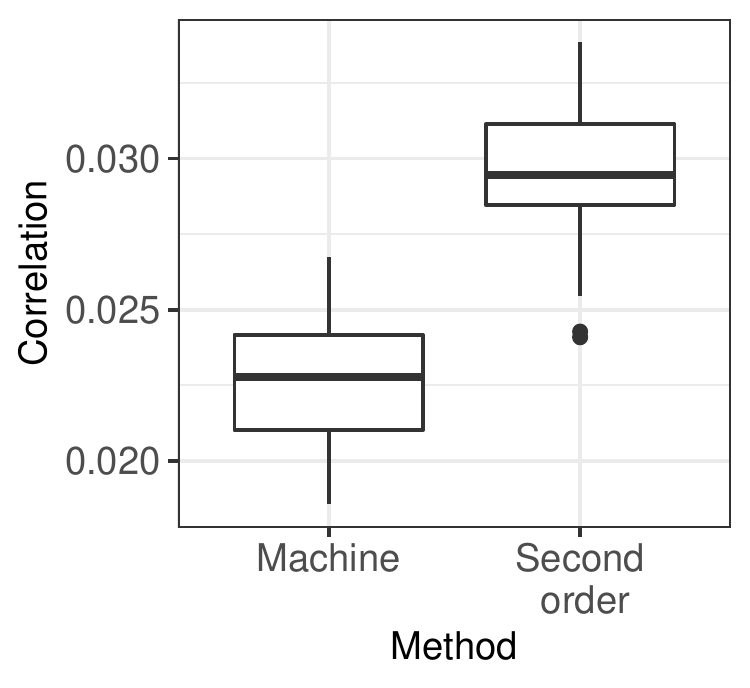}
  \caption{Boxplot comparing the average absolute pairwise correlation between variables and knockoffs for a sparse multivariate Gaussian distribution. The other details are as in Figure~\ref{fig:sim-gaussian-diagnostics-self}.} \label{fig:sim-sparse-diagnostics-self}
\end{figure}

Finally, a second example is presented in which second-order knockoffs do not control the false discovery rate. The distribution considered here involves variables that are weakly correlated but highly dependent. In particular, a random variable $\eta \in \RR$ is sampled from a standard normal distribution, while a random subset $A$ of size $|A|=L$ is independently chosen from $\{1,\ldots,p\}$. Then, for each $j \in \{1,\ldots,p\}$, the value of $X_j$ is given by
\begin{align*}
  X_j = \sqrt{\frac{{L \choose p}}{{L-1 \choose p-1}}} \cdot
  \begin{cases}
    \eta, & \text{if} \quad j \in A, \\
    0, & \text{otherwise}.
  \end{cases}
\end{align*}
The scaling factor ensures that each variable has unit variance.
In fact, the covariance matrix $\Sigma$ corresponding to this choice
of $P_X$ can easily be shown to be equal to
\begin{align*}
  \Sigma_{ij}
  & = \begin{cases}
    1, & \text{ if } i =j,  \\
    \frac{L-1}{p-1}, & \text{ otherwise. }
  \end{cases}
\end{align*}
Here, we choose $L=30$. Then, we perform the usual controlled numerical experiment on the deep machine trained with hyperparameters equal to $(\gamma,\lambda,\delta)=(1,0.1,1)$, using $m=200$ samples and setting $\alpha=0$ in \eqref{eq:knock-stats}. The hyperparameter $\lambda=0.1$ decreases the weight given to the empirical covariance matrix, as in the previous experiment, while $\delta=1$ ensures that the knockoffs are powerful. 
The performance of this machine is only compared to that of the second-order approximation. The results are shown in Figure~\ref{fig:sim-sparse-test}, while the goodness-of-fit diagnostics can be found in Figures \ref{fig:sim-sparse-diagnostics} and \ref{fig:sim-sparse-diagnostics-self}. We can see that the knockoffs generated by the machine are not exact; however, their approximation is more accurate than that of the second-order method. This improvement is also reflected in Figure~\ref{fig:sim-sparse-test}, illustrating that the deep machine leads to successful control of the false discovery rate, unlike the second-order knockoffs. A combination of careful parameter tuning of the loss function, different network design and a larger training set may even further improve quality.

\section{Application} \label{sec:application}
\subsection{Overview of the data} \label{sec:hiv-data}

We deploy the deep knockoff machine to a study of variations in drug
resistance among human immunodeficiency viruses of type I in order to
detect important mutations \cite{rhee2006genotypic}. We choose this
application mainly for its importance and because the data are freely
available from
\url{http://hivdb.stanford.edu/pages/published_analysis/genophenoPNAS2006/}. 
Moreover, an earlier release with fewer samples also appears in the first paper on knockoffs
\cite{barber2015}. It should be acknowledged that it
is not immediately clear whether the underlying assumptions of the
model-X settings are really satisfied. In particular, we do not know
how realistically the samples can be described as independent and
identically distributed pairs $(X,Y)$ drawn from some joint underlying
distribution. Rigorously validating these assumptions would require
expert domain knowledge and additional data. Therefore, we interpret
the analysis in this paper as an illustrative example of how deep
knockoff machines can be used in practice, without advancing any claim
of new scientific findings. In any case, it is encouraging to verify
that many of the mutations discovered by our method are already known
to be important, as discussed in Section~\ref{sec:hiv-res} and
Appendix \ref{sec:hiv-discoveries}.

For simplicity, we focus on analyzing the resistance to one protease
inhibitor drug, namely lopinavir. The response variable $Y^i$
represents the log-fold increase in resistance measured in the $i$th
virus. Having removed all samples containing missing values, we are
left with $n=1431$. Each of the $p=150$ binary features $X_j$
indicates the presence of a particular mutation. Half are chosen
because they are previously known to be associated with changes in the
drug resistance. The other half are chosen because they are the most
frequently occurring mutations. If multiple mutations occur at the
same position, the first two are treated as distinct while the others
are ignored. The variables are standardized to have zero mean and unit
variance, even though they have binary support.  The machine in
Section~\ref{sec:experiments-setup} is slightly modified by adding a
sigmoid activation function and an affine transformation on each
output node.  The hyperparameters in the loss function are
$(\gamma,\lambda,\delta)=(1,1,1)$.  The machine is trained after
$T = 5 \times 10^4$ gradients steps and a learning rate $\mu = 0.01$.

The strategy adopted for the analysis of these data is different from that
described in the simulations of Section \ref{sec:experiments}. 
A deep knockoff machine is trained on the 150 mutation features 
corresponding to all 1431 subjects.
Since the data is limited, we fit the machine on the same samples for which we
need to generate the knockoff copies to perform variable selection.
Therefore, it is possible that some overfitting will occur.
In other words, even though the machine thus obtained may not be very 
accurate on new observations of $X$, the knockoffs produced on the training set
will be nearly indistinguishable upon a finite-sample swap with the original variables.
Overfitting knockoffs has been empirically observed to lead to a loss of power at worst, while the control of the type-I errors typically remains intact \cite{candes2016panning, sesia2017gene, lu2018deeppink}. This claim is confirmed by the results of the numerical experiments presented below, although future research should investigate a theoretical explanation of this phenomenon. For now, we accept this limitation and proceed by verifying that the machine works for our purposes. 

\subsection{Numerical experiments with real variables} \label{sec:hiv-sim}

The numerical experiments described in Section~\ref{sec:experiments}
are carried out using artificial response variables simulated from a
known conditional linear model with 30 nonzero coefficients. Since the
true population distribution of the mutations is unknown, new
observations cannot be drawn from $P_X$. Instead, each experiment is
carried out on a randomly chosen subset of size $m < n$ of the
original data. Two different values of $m \in \{200,300\}$ are
considered, as discussed below. Variable selection is based on the
same importance statistics of Section~\ref{sec:experiments-setup},
setting $\alpha = 0.1$ in \eqref{eq:knock-stats}, and applying the
knockoff filter to control the false discovery rate at the nominal
level $q=0.1$. As the ground truth is known, the number of true and
false discoveries can be evaluated. The natural benchmark for our
machine is the second-order method in \cite{candes2016panning}.  The
empirical covariance matrix for the latter is evaluated on the full
data, in order to make a fair comparison with the deep machine.  Even
though the fixed-X knockoffs in \cite{barber2015} could in principle
be used in the case where $m \geq 2p$, we have observed that they are
severely underpowered in this simulation. In fact, the features
exhibit strong correlations and the empirical covariance matrix in
subsets of the data of size $m=300$ is frequently singular. Therefore,
fixed-X knockoffs are not ideally suited for this numerical experiment
and a plot of their performance is omitted.

\begin{figure}[!htb]
    \centering
    \begin{subfigure}[t]{0.49\textwidth}
        \centering
        \includegraphics[width=0.8\textwidth]{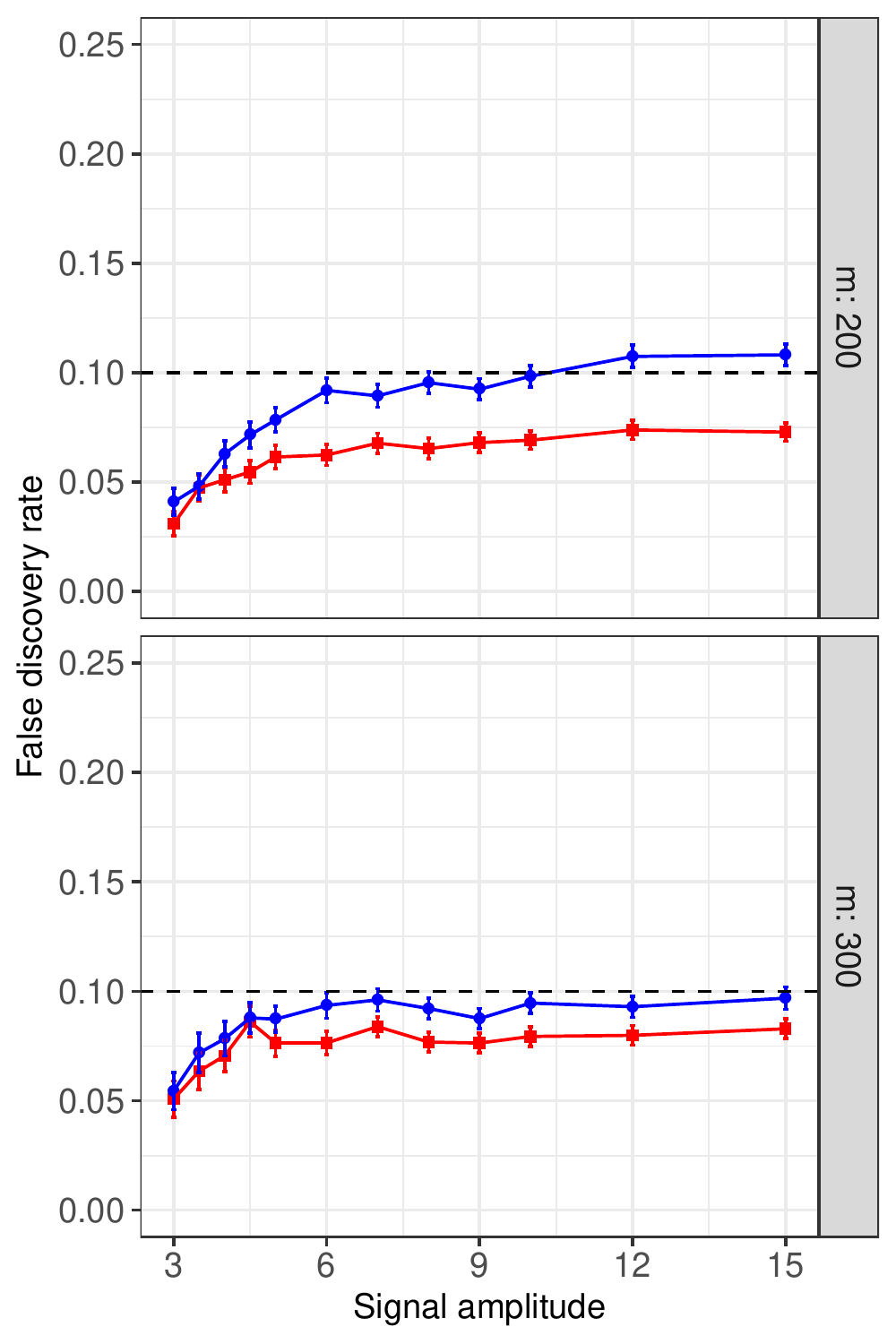}
        \caption{}  \label{fig:hiv-sim-fdr}
    \end{subfigure}
    \begin{subfigure}[t]{0.49\textwidth}
        \centering
        \includegraphics[width=0.8\textwidth]{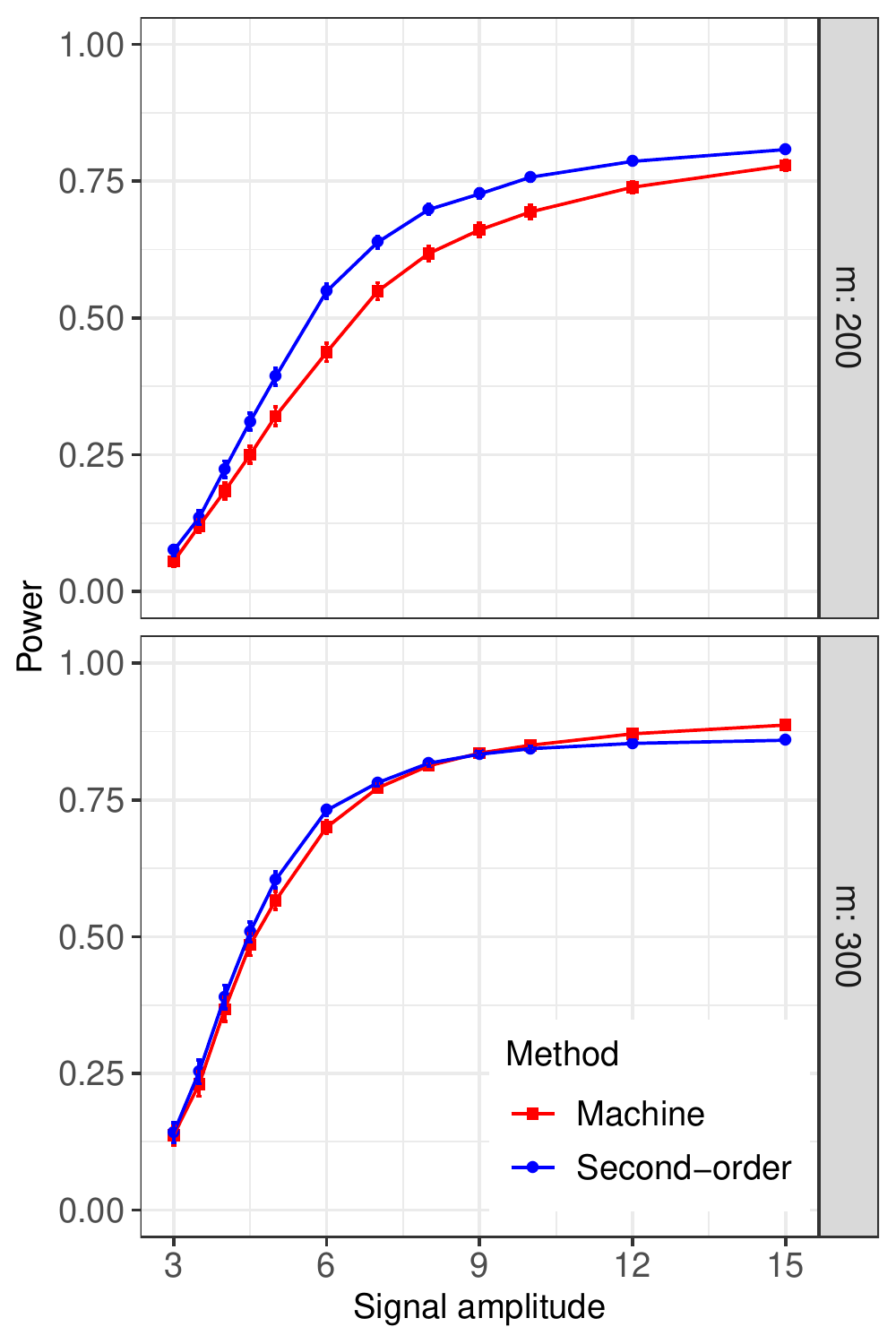}
        \caption{}  \label{fig:hiv-sim-pow}
    \end{subfigure}
  \caption{Numerical experiment with real human immunodeficiency virus mutation features and simulated response. The performance of the deep machine (red) is compared to that of second-order knockoffs (blue). The false discovery rate (a) and the power (b) are averaged over 1000 replications. Each replication is performed on a random subset of the original data containing $m=200$ (top) or 300 (bottom) observations chosen without replacement.} \label{fig:hiv-sim}
\end{figure}

The results are shown in Figure~\ref{fig:hiv-sim} as a function of the signal amplitude. The deep machine successfully controls the false discovery rate, while the second-order method is slightly too liberal. Deep knockoffs often lead to more true discoveries than the second-order approximation, while making fewer mistakes. We believe that the model-X knockoff constructions applied here are overfitting, although the deep machine is more effective for variable selection within the dataset. Further insight may be provided by the goodness-of-fit diagnostics in Section~\ref{sec:tuning}; however, these would require access to additional independent samples from the same population, which we unfortunately lack. As a partial solution, one  could try to split the data, even though it is not clear whether the cost would be justified, since the diagnostics will not be very powerful when evaluated on small samples.

\begin{figure}[!htb]
    \centering
    \begin{subfigure}[t]{0.49\textwidth}
        \centering
        \includegraphics[width=0.8\textwidth]{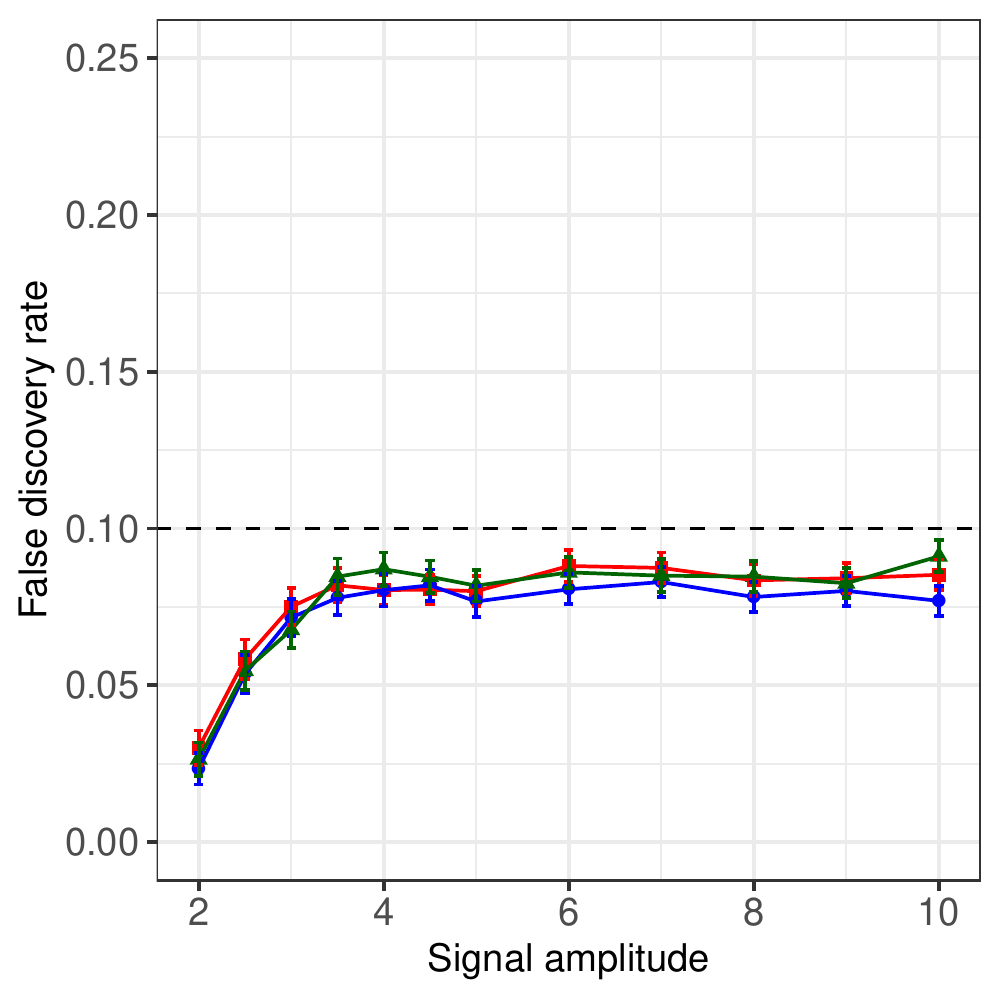}
        \caption{}  \label{fig:hiv-sim-full-fdr}
    \end{subfigure}
    \begin{subfigure}[t]{0.49\textwidth}
        \centering
        \includegraphics[width=0.8\textwidth]{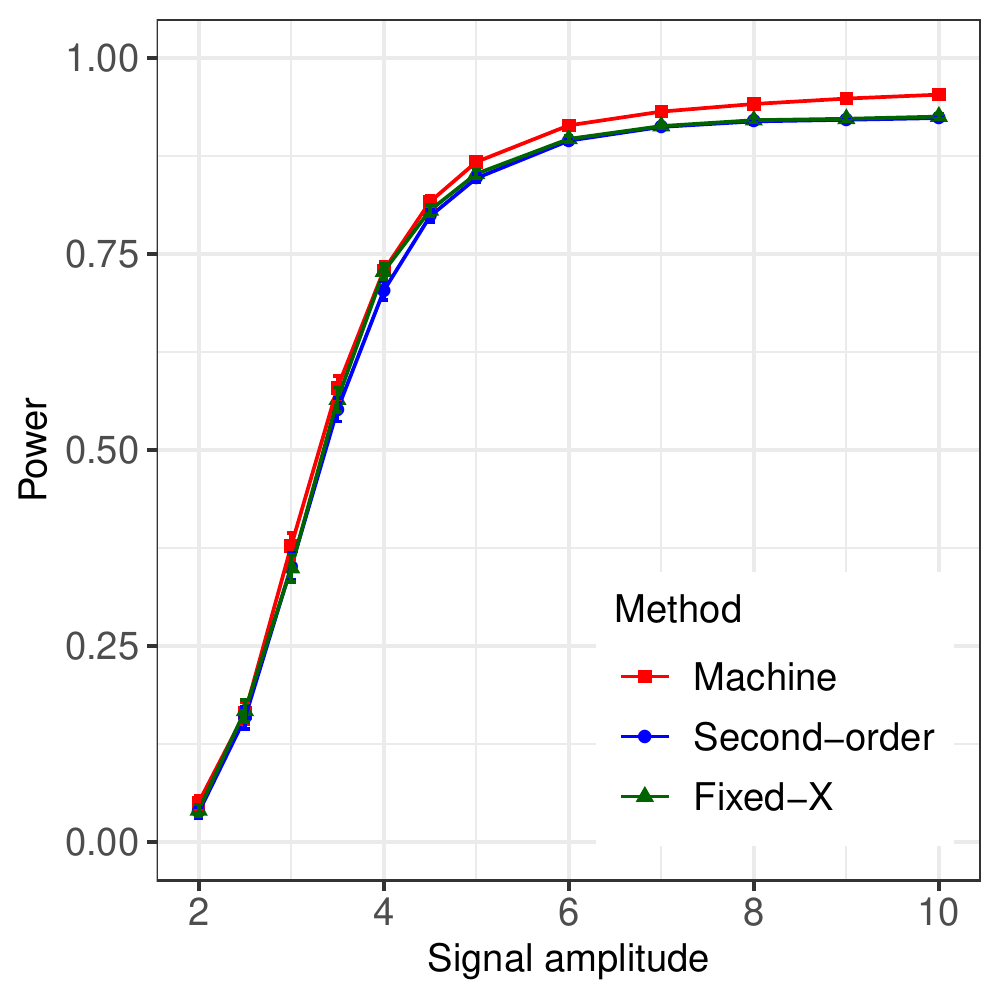}
        \caption{}  \label{fig:hiv-sim-full-pow}
    \end{subfigure}
  \caption{Numerical experiment with real human immunodeficiency virus mutation features and simulated response. The performance of the deep machine (red) is compared to that of second-order knockoffs (blue) and fixed-X knockoff (green). The false discovery rate (a) and the power (b) are averaged over 1000 replications. Each replication is performed on the fixed original features $\X$.} \label{fig:hiv-sim-full}
\end{figure}

Our numerical experiments can be slightly altered to see what happens
when we hold $\X \in \reals^{1431 \times 150}$ constant and simulate a
response variable for each observation.  In theory, model-X knockoffs
may not control the false discovery rate conditional on $\X$.
However, it can be informative to apply and compare in this context
the procedures described above.  Since $n$ is much greater than $p$
and $\X$ is fixed, fixed-X knockoffs are a reasonable alternative to
the deep machine and the second-order method. The results
corresponding to the three competing approaches averaged over 1000
replications are shown in Figure \ref{fig:hiv-sim-full} as a function
of the signal amplitude.  It is reassuring to observe that the
second-order and the fixed-X knockoffs appear to control the false
discovery rate and achieve similar power, while the deep machine
outperforms both.

\subsection{Results} \label{sec:hiv-res}

\begin{figure}[!htb]
  \centering
    \includegraphics[width=0.4\textwidth]{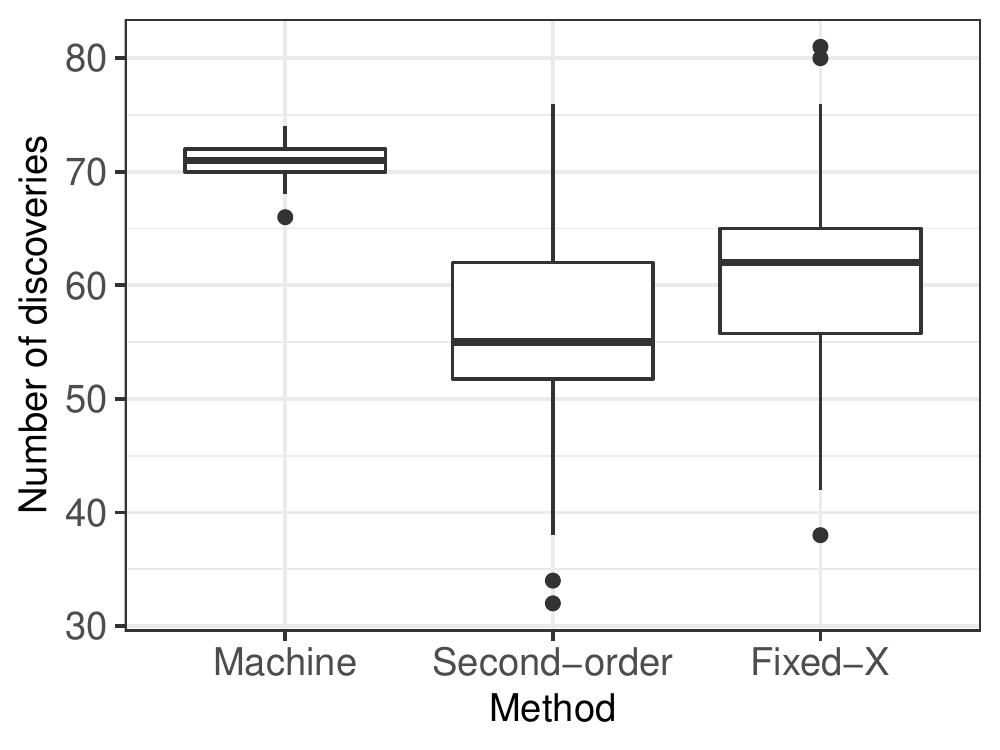}
  \caption{Boxplot of the number of drug-resistance mutations in the human immunodeficiency virus discovered using different knockoff generation methods. The variability in the results corresponds to 100 independent samples of the knockoff copies.}\label{fig:hiv-discoveries}
\end{figure}

Finally, the knockoffs generated by the machine trained in
Section~\ref{sec:hiv-sim} are used to select important features that
contribute to explaining changes in the drug resistance of the
viruses. The knockoff filter is applied using the same importance
statistics as above, setting $\alpha = 0.1$ in
\eqref{eq:knock-stats}. The nominal false discovery rate is
$q=0.1$. In order to investigate the stability of the findings
obtained with this machine, the variable selection procedure is
repeated 100 times, starting from a new independent realization of the
knockoffs conditional on the data. The distribution of the number of
discoveries on this dataset is displayed in
Figure~\ref{fig:hiv-discoveries}, along with the analogous quantity
corresponding to second-order knockoffs \cite{candes2016panning} and
the randomized version of the fixed-X knockoffs \cite{barber2015}.
The results indicate that the deep machine leads to more discoveries
than the alternative approaches. This is in line with the numerical
experiments presented above. It should not be surprising that fixed-X
knockoffs perform better here than in Section~\ref{sec:hiv-sim}
because the sample size is much larger. It is interesting that the
selections made with our machine are quite stable upon resampling of
$\tilde{X} \mid X$, unlike those of other methods. This potentially
significant advantage of deep knockoff machines should be investigated
more rigorously in future work.  The list of discovered mutations is
in large part consistent with the prior knowledge on their importance,
and it is shown in Appendix \ref{sec:hiv-discoveries}. In fact,
according to the database on
\url{https://hivdb.stanford.edu/dr-summary/comments/PI/}, many of our
findings have been previously reported to have a major or accessory
effect on changes in drug resistance.


\section{Discussion} \label{sec:conclusion}

\subsection{Summary}

The deep machines presented in this paper extend the knockoff method to a vast range of problems. The idea of sampling knockoff copies by matching higher moments is a natural generalization of the existing second-order approximation; however, the inherent difficulties of this approach have prompted us to exploit the powerful new methods of deep learning. The numerical experiments and the data analysis described in this paper can be reproduced on a single graphics processing unit within a few hours. We believe that the computational cost can be decreased as more experience is acquired, and applications on a larger scale should be pursued. The extensive numerical experiments show that our solution can match the performance of the available exact knockoff constructions for several data distributions, and greatly outperform the previous approximations in more complex cases. The diagnostics computed on independent test data confirm that the deep machines are correctly learning to generate valid knockoffs, without relying on prior knowledge. The theoretical results contribute to providing a principled basis for our approach. The outcomes of the data analysis are also encouraging and motivate further applications.

There is a subtle but meaningful difference between the perspective taken by the existing theory of model-X knockoffs and the common practice on real data. In principle, finite-sample control of the false discovery rate is guaranteed when the knockoff copies are constructed with respect to the true $P_X$. The work of \cite{barber2018robust} precisely quantifies the extent of the worst-case deviations that may occur when a fixed and misspecified distribution $Q_X$ is used instead of $P_X$. However, knockoffs are often constructed using an estimated $\hat{P}_X$ obtained from the same samples used for variable selection, as discussed in Section~\ref{sec:application}. The interesting empirical observation is that when $\hat{P}_X$ overfits the training samples, knockoffs typically become more conservative rather than too liberal. To the best of our knowledge, this phenomenon is still lacking a rigorous explanation. In any case, the numerical simulations of Section~\ref{sec:experiments} show that deep knockoff machines can learn how to generate valid knockoffs. In conclusion, we believe that this work is a valuable contribution because it allows the rich framework of knockoffs to be applied in very general settings. In fact, given sufficient data and adequate computing resources, deep knockoff machines can be trained on virtually any kind of features.

\subsection{Our experience with other machines}

This work has been primarily driven by the need to develop an effective and principled tool for the analysis of complex datasets. Generative moment matching networks are not the only technique that we considered. In fact, the worst-case perspective in the robustness theory may suggest an approach based on generative adversarial networks. We were discouraged from following that route by the complications of simultaneously fitting a generator and a discriminator. Moreover, we were mainly driven by a desire to seek a simple and practical solution, preferably building upon the well established second-order approximation. Our considerable efforts in the attempt to generate good knockoffs with a variational autoencoder could not overcome the serious limitations in power that we observed. An alternative solution was also explored, relying on deep Boltzmann machines to learn a suitably exchangeable joint distribution of $(X,\tilde{X})$ that would be consistent with the observed data. However, the computational challenges resulting from this fundamentally more difficult stochastic optimization problem eventually convinced us to search for a better path. Finally, we have found that generative moment matching networks lead to deep knockoff machines that are very effective and elegantly fit within the existing literature.

\subsection{Future work}

There are several paths open for future research. For example,
variations of our machines could be based on different knockoff
scoring functions or different regularization penalties. The deep
machines described in this paper take a completely agnostic view of
the data distribution, but there are many applications in which some
prior knowledge of the structure of the variables is
available. Exploiting it could greatly improve the computational and
statistical efficiency of our method. An example arises from
genome-wide association studies, where the features are naturally
arranged in a sequential order and exhibit local dependencies that can
be well described by a hidden Markov model. It may be interesting to
develop deep knockoff machine specialized for this setting and to
compare it with the procedure of \cite{sesia2017gene} on a large
scale. A different project could involve the extension of our toolbox
of diagnostics and a systematic study of their relative strengths. An
extension of the theoretical results in \cite{barber2018robust} may
also be valuable. Since alternative knockoff constructions based on
different deep learning techniques have been independently proposed in
parallel to the writing of this paper \cite{anonymous2019knockoffgan,
  liu2018auto}, it is also up to future research to extensively
compare their empirical performance.


\subsection*{Acknowledgements}
E. C. was partially supported by the Office of Naval Research (ONR)
under grant N00014-16- 1-2712, by the Army Research Office (ARO) under
grant W911NF-17-1-0304, by the Math + X award from the Simons
Foundation and by a generous gift from TwoSigma. Y. R. was supported
by the same Math + X award and ARO grant. Y.~R.~also thanks the Zuckerman
Institute, ISEF Foundation and the Viterbi Fellowship, Technion, for
supporting this research. M.S. was supported by the same Math + X
award and ONR grant. We thank Stephen Bates, Nikolaos Ignatiadis and
Eugene Katsevich for their insightful comments on an earlier draft of
this paper.




\bibliographystyle{ieeetr}
\bibliography{bibliography}

\begin{appendices}
\section{Proofs} \label{sec:proofs}
\begin{proof}[Proof of Theorem \ref{thm:machine-loss}]
  Recall that $J_{ \text{MMD}}$ is defined as:
  \begin{align*}
    \EE \left[ J_{ \text{MMD}}(\X, \tilde{\X})\right]
    & =  \EE \left\{ \widehat{\mathcal{D}}_{ \text{MMD}} \left[ (\X',\tilde{\X}'), (\tilde{\X}'', \X'')\right]  \right\} +
      \EE \left\{ \widehat{\mathcal{D}}_{ \text{MMD}} \left[ (\X',\tilde{\X}'), (\X'',\tilde{\X}'')_{\text{swap}(S)} \right] \right\} \\
    & =  \EE \left\{ \widehat{\mathcal{D}}_{ \text{MMD}} \left[ (\X',\tilde{\X}'), (\tilde{\X}'', \X'')\right]  \right\} +
      \EE \left\{ \EE \left[ \widehat{\mathcal{D}}_{ \text{MMD}} \left[ (\X',\tilde{\X}'), (\X'',\tilde{\X}'')_{\text{swap}(S)} \right] \mid S \right] \right\}.
  \end{align*}
  The expectation is taken with respect to the random swap $S$, the data $\X$, its partition into $\X',\X''$ and the noise in the machine that produces $\tilde{\X}', \tilde{\X}''$. Since we know from \cite{gretton2012kernel} that
  $\widehat{\mathcal{D}}_{ \text{MMD}}$ is an unbiased estimator of
  $\mathcal{D}_{ \text{MMD}}$ and that the latter is a non-negative
  quantity, it follows immediately that
  $\EE \left[ J_{ \text{MMD}}(\X, \tilde{\X})\right] \geq 0$. 
Furthermore, the samples in $\X$ are independent and identically distributed and the partition is randomly chosen. Therefore, it follows that
  \begin{align} \label{eq:proof-JMMD}
    \begin{split}
    \EE \left[ J_{ \text{MMD}}(\X, \tilde{\X})\right]
    & = \mathcal{D}_{ \text{MMD}} \left[ P_{(X',\tilde{X}')}, P_{(\tilde{X}'', X'')}\right] + 
      \EE \left\{ \mathcal{D}_{ \text{MMD}} \left[ P_{(X',\tilde{X}')}, P_{(X'',\tilde{X}'')_{\text{swap}(S)}} \right] \right\} \\
    & = \mathcal{D}_{ \text{MMD}} \left[ P_{(X,\tilde{X})}, P_{(\tilde{X}, X)}\right] + 
      \EE \left\{ \mathcal{D}_{ \text{MMD}} \left[ P_{(X,\tilde{X})}, P_{(X,\tilde{X})_{\text{swap}(S)}} \right] \right\}.
    \end{split}
  \end{align}
  Above, the remaining expectation is taken over the random swap
  $S$. We know that the first term is equal to zero if and only if
  $(\tilde{X},X)$ has the same distribution as
  $(X,\tilde{X})$. Moreover, if
  $(X,\tilde{X})_{\text{swap}(j)} \overset{d}{=} (X,\tilde{X})$ for
  all $j \in \{1,\ldots,p\}$, it follows that
  $(X,\tilde{X})_{\text{swap}(S)} \overset{d}{=} (X,\tilde{X})$ for
  all $S \subset \{1,\ldots,p\}$ (see \cite{candes2016panning}) and
  the second term is zero. Conversely, assuming that the second term
  in \eqref{eq:proof-JMMD} is equal to zero implies that
  $(X,\tilde{X})_{\text{swap}(S)} \overset{d}{=} (X,\tilde{X})$ for
  all $S \subset \{1,\ldots,p\}$. The last conclusion holds because
  any subset $S$ has a positive probability of being chosen and the
  maximum mean discrepancy between any two distributions is always
  positive unless they are equal, in which case it vanishes.

\end{proof}

\begin{proof}[Proof of Theorem \ref{thm:machine-convergence}]
The strategy is inspired by Theorem 2.1 and Corollary 2.2 in \cite{ghadimi2013stochastic}, as well as Theorem 4.1 and Remark 4.2.1 in \cite{sanjabi2018solving}. While the convergence result in \cite{ghadimi2013stochastic} refers to a modified version of stochastic gradient descent in which the number of gradient steps is random, we consider a fixed number of steps. The approach in \cite{sanjabi2018solving} is closer to ours. 

First, it follows from the gradient update rule and a first-order expansion that
\begin{align*}
J_{\theta_{t+1}} & \leq J_{\theta_t} + \langle \nabla J_{\theta_t}, \theta_{t+1} - \theta_t \rangle + \frac{L}{2}\mu^2 \|g_t\|^2_2.
\end{align*}
Defining $\delta_t = g_t - \nabla J_{\theta_t}$, the relation $-\mu g_t = \theta_{t+1} - \theta_t$ can be used to manipulate the above inequality as in \cite{ghadimi2013stochastic}:
\begin{align*} 
J_{\theta_{t+1}} & \leq J_{\theta_t} - \mu \langle \nabla J_{\theta_t}, g_{t} \rangle + \frac{L}{2}\mu^2 \|g_t\|^2_2 \\
& = J_{\theta_t} - \mu \| \nabla J_{\theta_t} \|_2^2  - \mu \langle \nabla J_{\theta_t}, \delta_{t} \rangle + \frac{L}{2}\mu^2 \left( \|\nabla J_{\theta_t}\|^2_2 + 2 \langle \nabla J_{\theta_t}, \delta_{t} \rangle + \|\delta_{t}\|_2^2\right) \\
& = J_{\theta_t} - \left(\mu - \frac{L}{2}\mu^2\right) \| \nabla J_{\theta_t} \|_2^2  - \left(\mu - L \mu^2 \right)\langle \nabla J_{\theta_t}, \delta_{t} \rangle + \frac{L}{2}\mu^2 \|\delta_t\|^2_2.
\end{align*}
Summing the above inequalities over $t=1,\ldots,T$ we get
\begin{align}
	\begin{split}\label{Eq:normGrad}
	\left( \mu - \frac{L}{2}\mu^2\right)\sum_{t=1}^T\|\nabla J_{\theta_t}\|^2_2 
	& \leq J_{\theta_1} - J_{\theta_{T+1}} - \left(\mu - L \mu^2\right) \sum_{t=1}^T  \langle \nabla J_{\theta_t}, \delta_t \rangle + \frac{\mu^2L}{2}\sum_{t=1}^T \| \delta_t \|^2_2  \\
	& \leq J_{\theta_1} - J^* - \left(\mu - L \mu^2\right) \sum_{t=1}^T  \langle \nabla J_{\theta_t}, \delta_t \rangle + \frac{\mu^2L}{2}\sum_{t=1}^T \| \delta_t \|^2_2.
	\end{split}
\end{align}
We now take the expectation of \eqref{Eq:normGrad} on both sides, conditional on $\zeta_1$. Since the estimated gradients $g_t$ are unbiased, i.e.~$\Ec{g_t}{\zeta_t} = \nabla J_{\theta_t}$, we have 
\begin{align*}
  \Ec{\langle \nabla J_{\theta_t}, \delta_t \rangle }{\zeta_1} 
  & = \Ec{ \Ec{\langle \nabla J_{\theta_t}, \delta_t \rangle}{\zeta_1, \zeta_t} }{\zeta_1}\\
  & = \Ec{ \Ec{\langle \nabla J_{\theta_t}, \delta_t \rangle}{\zeta_t} }{\zeta_1}\\
  & = \Ec{ \langle \nabla J_{\theta_t}, \Ec{\delta_t}{\zeta_t} \rangle }{\zeta_1}\\
  & = \Ec{ \langle \nabla J_{\theta_t}, 0 \rangle }{\zeta_1} = 0.
\end{align*}
Substituting this result into \eqref{Eq:normGrad}, and using the assumption that $\Ec{\| \delta_t\|^2_2}{\zeta_t} \leq \sigma^2$, leads to
\begin{align*}
\left( \mu - \frac{L}{2}\mu^2\right) \sum_{t=1}^T \Ec{\|\nabla J_{\theta_t}\|^2_2}{\zeta_1} & \leq J_{\theta_1} - J^* + \frac{\mu^2L}{2} T \sigma^2.  
\end{align*}
Finally, multiplying both sides by $2 / [LT (2\mu  - L\mu^2)]$ results in
\begin{align*}
\frac{1}{TL}\sum_{t=1}^T \Ec{\|\nabla J_{\theta_t}\|^2_2}{\zeta_1} 
  & \leq \frac{2(J_{\theta_1} - J^*)}{TL\left( 2\mu - L\mu^2\right)}  + 
    \frac{\sigma^2 \mu}{ 2 - L\mu} \\
  & \leq \frac{\Delta}{T \left( 2\mu - L\mu^2\right)}  + 
    \frac{\sigma^2 \mu}{2 - L\mu}.
\end{align*}
The choice of $\mu = \min\left\{\frac{1}{L}, \frac{\mu_0}{\sigma \sqrt{T} } \right\}$ allows us to conclude that
\begin{align*}
\frac{1}{TL}\sum_{t=1}^T \Ec{\|\nabla J_{\theta_t}\|^2_2 }{\zeta_1}
  & \leq \frac{L\Delta}{T} + \left(\mu_0 + \frac{\Delta}{\mu_0}\right)\frac{\sigma}{\sqrt{T}}.
\end{align*} 
\end{proof}

\section{Table of discoveries for the HIV dataset} \label{sec:hiv-discoveries}

\begin{longtable}[t]{ccccc}
\caption{\label{tab:}\label{tab:hiv-discoveries}List of drug-resistance mutations discovered using a deep knockoff machine, compared to the findings obtained with second-order and fixed-X knockoffs. The annotation refers to existing knowledge available on the importance of each mutation.}\\
\toprule
Mutation & Annotation & Machine & Second-order & Fixed-X\\
\midrule
46L & Major & 100 & 100 & 100\\
47A & Major & 100 & 100 & 100\\
47V & Major & 100 & 100 & 100\\
48V & Major & 100 & 100 & 100\\
50L & Major & 100 & 100 & 100\\
50V & Major & 100 & 100 & 100\\
54A & Major & 100 & 100 & 100\\
54V & Major & 100 & 100 & 100\\
76V & Major & 100 & 100 & 100\\
82F & Major & 100 & 100 & 100\\
82S & Major & 100 & 100 & 100\\
82T & Major & 100 & 100 & 100\\
84A & Major & 100 & 100 & 100\\
54S & Major & 100 & 99 & 100\\
54L & Major & 100 & 98 & 99\\
82C & Major & 100 & 97 & 100\\
54M & Major & 100 & 96 & 95\\
48M & Major & 100 & 90 & 94\\
32I & Major & 100 & 82 & 81\\
48L & Major & 100 & 73 & 72\\
30N & Major & 100 & 56 & 62\\
48Q & Major & 100 & 18 & 26\\
46I & Major & 100 & 0 & 0\\
82A & Major & 100 & 0 & 0\\
84V & Major & 100 & 0 & 0\\
90M & Major & 100 & 0 & 0\\
10F & Accessory & 100 & 100 & 100\\
20T & Accessory & 100 & 100 & 100\\
24F & Accessory & 100 & 100 & 100\\
24I & Accessory & 100 & 100 & 100\\
43T & Accessory & 100 & 100 & 100\\
73C & Accessory & 100 & 93 & 97\\
73S & Accessory & 100 & 88 & 95\\
73T & Accessory & 100 & 85 & 79\\
58E & Accessory & 100 & 73 & 74\\
53L & Accessory & 100 & 63 & 74\\
23I & Accessory & 100 & 59 & 71\\
33F & Accessory & 100 & 0 & 0\\
88D & Accessory & 100 & 0 & 0\\
10V & Other & 100 & 99 & 99\\
20I & Other & 100 & 70 & 81\\
71I & Other & 100 & 41 & 46\\
10I & Other & 100 & 0 & 0\\
71V & Other & 100 & 0 & 0\\
16A & NA & 100 & 100 & 100\\
72M & NA & 100 & 99 & 98\\
89I & NA & 100 & 95 & 98\\
67F & NA & 100 & 94 & 96\\
57K & NA & 100 & 94 & 92\\
35N & NA & 100 & 93 & 91\\
77I & NA & 100 & 92 & 96\\
95F & NA & 100 & 92 & 96\\
69K & NA & 100 & 84 & 92\\
64L & NA & 100 & 84 & 84\\
37D & NA & 100 & 79 & 90\\
22V & NA & 100 & 79 & 82\\
92K & NA & 100 & 75 & 94\\
93L & NA & 100 & 65 & 72\\
67E & NA & 100 & 59 & 64\\
91S & NA & 100 & 58 & 63\\
66V & NA & 100 & 57 & 79\\
12A & NA & 100 & 41 & 56\\
72R & NA & 100 & 38 & 45\\
37C & NA & 100 & 28 & 36\\
36I & NA & 100 & 0 & 0\\
63P & NA & 100 & 0 & 0\\
14R & NA & 99 & 19 & 21\\
73A & Accessory & 98 & 18 & 21\\
20M & Other & 94 & 3 & 12\\
12S & NA & 86 & 22 & 34\\
84C & Major & 57 & 40 & 49\\
74S & Other & 42 & 31 & 34\\
45R & NA & 24 & 72 & 93\\
43R & NA & 6 & 18 & 24\\
\bottomrule
\end{longtable}

\end{appendices}

\end{document}